\documentclass[preprint]{aastex}

\newcommand\cs{c_s}
\newcommand\dmax{\rho_{\rm max}}
\newcommand\dmid{\rho_{00}}
\newcommand\vA{v_{\rm A}}
\newcommand\gext{g_{\rm ext}}
\newcommand\geff{g_{\rm eff}}
\newcommand\kx{k_x}
\newcommand\ky{k_y}
\newcommand\kmax{k_{\rm max}}

\newcommand\Msun{{\rm\,M_\odot}}

\newcommand\torb{t_{\rm orb} }
\newcommand\tsh{t_{\rm sh} }
\newcommand\tParker{t_{\rm Parker} }
\newcommand\tgrav{t_{\rm grav} }
\newcommand\tswing{t_{\rm sw} }
\newcommand\pc{{\rm\,pc}}
\newcommand\kpc{{\rm kpc}}
\newcommand\simgt{\lower.5ex\hbox{$\; \buildrel > \over \sim \;$}}
\newcommand\simlt{\lower.5ex\hbox{$\; \buildrel < \over \sim \;$}}

\shortauthors{Kim, Ostriker, \& Stone}
\shorttitle{Parker, Magneto-Jeans, and Swing Instabilities}

\begin{document}

\pagebreak
\title{Three-Dimensional Simulations of Parker, Magneto-Jeans, and
Swing Instabilities in Shearing Galactic Gas Disks}

\author{Woong-Tae Kim, Eve C. Ostriker, and James M. Stone}
\affil{Department of Astronomy, University of Maryland \\
College Park, MD 20742-2421}

\email{kimwt@astro.umd.edu, ostriker@astro.umd.edu, jstone@astro.umd.edu}

\slugcomment{Accepted for publication in the Astrophysical Journal
\hspace{0.98cm}}

\begin{abstract}
Various instabilities have been proposed as candidates to prompt
the condensation of giant, star-forming cloud complexes from
the diffuse interstellar medium. Here,
we use three-dimensional ideal MHD simulations to investigate nonlinear
development of the Parker, magneto-Jeans (MJI),  and swing mechanisms
in galactic disk models.
The disk models are local, isothermal, and begin from a vertically-stratified 
magnetohydrostatic equilibrium state with both gaseous and stellar gravity.
We allow for a range of surface densities and
rotational shear profiles, as well as unmagnetized control models.
We first construct axisymmetric equilibria and examine their stability. 
Finite disk thickness reduces the critical
Toomre stability parameter below unity; we find
$Q_c\sim0.75$, 0.72, and 0.57 for zero, sub-equipartition, and
equipartition magnetic field cases, respectively.
We then pursue fully three-dimensional models. 
In non-self-gravitating cases, the peak mid-disk density enhancement 
from the ``pure'' Parker instability is less a factor of two. 
The dominant growing modes have radial wavelengths $\lambda_x$
comparable to the disk scale height $H$, much shorter than the 
azimuthal wavelength ($\lambda_y\sim10-20H$).
Shearing disks, being more favorable to midplane-symmetric modes, 
have somewhat different late-time magnetic field profiles from 
nonshearing disks, but otherwise saturated states are similar.
Late-time velocity fluctuations at 10\% of the sound speed persist,
but no characteristic structural signatures of Parker modes 
remain in the new quasi-static equilibria.
In self-gravitating cases, the development of density structure is 
qualitatively similar to our previous results from thin-disk simulations.
The Parker instability, although it may help seed structure or
tip the balance under marginal 
conditions, appears to play a secondary role -- not affecting,
for example, the sizes or spacings of the bound structures that form.
In shearing disks with $Q$ less than a threshold level $\approx 1$, swing 
amplification can produce bound clouds of a few times the local Jeans mass.
The most powerful cloud-condensing mechanism,
requiring low-shear conditions as occur in spiral arms or galactic 
centers, appears to be the MJI. 
In thick disks, the MJI occurs for $\lambda_y\simgt 2\pi H$.
Our simulations show that condensations of a local Jeans mass 
($\simlt3\times 10^7\,\Msun$) grow very rapidly, supporting the idea that MJI
is at least partly responsible for the formation of bound
cloud complexes in spiral galaxies.
\end{abstract}

%for ApJ keywords

\section{Introduction}

Nearly half of the interstellar medium (ISM) in the inner part of the
Milky Way is estimated to be in the molecular component 
(e.g., \citealt{dam93}), with the largest portion of the molecular ISM 
in giant molecular clouds or cloud complexes (GMCs) of
masses $\sim10^4-10^6\,\Msun$ \citep{dam87,sol87}, where most of
star formation occurs (e.g., \citealt{wil97}). 
GMCs are generally turbulent, self-gravitating,
and magnetized (e.g., \citealt{bli93,cru99}).
In external galaxies, they often appear in clusters, forming giant molecular
associations (GMAs) (e.g., \citealt{gra87, vog88, ran90, ran93, sak99}),
or within the boundaries of \ion{H}{1} superclouds (e.g., \citealt{elm95}).
These giant clouds in spiral galaxies are mostly associated with
spiral arms, but they sometimes appear in the interarm regions 
(e.g., \citealt{sol85,ran93,kenney97,hey98}).
By regulating the rate and mode of star formation in disk galaxies,
GMCs play a fundamental role in
controlling galactic evolution and may be a key to understanding
the nature of the Hubble sequence.
But how GMCs form is still not clearly understood.

Traditionally, proposed mechanisms for giant cloud formation fall into 
two categories: stochastic coagulation of smaller clouds
(following \citealt{oor54}; e.g., \citealt{kwa79}) or collective effects 
involving instabilities (see e.g., \citealt{elm96}). 
Existing work suggests that collisional agglomeration may proceed too slowly
(e.g., \citealt{sco79,elm90,elm95}).
Collisions in most cases result in disruption rather than merger
(e.g., \citealt{lat85,kim99,kle01}); the
available mass in smaller clouds is not enough to form
GMCs through collisional build-up \citep{hey98,bli99};
collisional agglomeration would lead to magnetically
supercritical clouds and active star formation before
GMC-scale masses are reached \citep{ost99,ost01};
GMC chemistry may be inconsistent with an extended build-up phase 
(e.g., \citealt{vandis93}).
All of these and other difficulties point to the need for collective 
effects \citep{bli80}.
This second means of building GMCs and GMAs involves large-scale dynamical
instabilities that coherently amass material over scales large
compared to the sizes of diffuse atomic clouds
(see e.g., \citealt{elm95}).  The small-scale
structure of the medium plays a secondary role, with the modal
gathering of material operating collectively on the whole
distribution.  Coherently-growing unstable modes that produce
condensations depend on various effects, including self-gravity,
the Coriolis force, sheared azimuthal motion, magnetic tension, and
magnetic buoyancy, all of which have been shown to interact in unstable
fashion.\footnote{Thermal instabilities involving changes in the 
microphysical kinetic temperature are important in forming cold 
clouds on small scales (\citealt{fie65}; see also e.g., \citealt{hen99}
and \citealt{bur00} for recent simulations); 
if dissipation of kinetic energy in collisions of cloudlets acts 
effectively like microscopic cooling for the turbulent ``cloud fluid'', 
then analogous mechanisms may also be 
important on large scales (e.g., \citealt{elm89, elm91}).}

The two most well-known condensation instabilities are the Parker
instability \citep{par66}, in which buoyancy causes dominantly
in-plane magnetic fields to buckle, with matter collecting in valleys;
and the swing amplifier, in which epicyclic motion conspires with
galactic shear to enable self-gravitating enhancement of overdense
regions \citep{gol65b,too81}.  Another
condensation instability that has been shown to be efficient in
low-shear, magnetized regions is the magneto-Jeans instability (MJI), in
which magnetic tension of in-plane fields counters the stabilizing
effect of the Coriolis force, allowing self-gravitating contraction of
overdense regions \citep{elm87,kim01}. 
Because the growth rates of Parker and MJI
modes increase with magnetic field strength and density, the compression of
the ISM induced by passage through stellar spiral arms may be
important in triggering GMC and GMA formation 
(e.g., \citealt{mou_etal74,bli80}).
Observed distributions of \ion{H}{2} regions and OB star complexes 
in a ``beads on a string'' pattern along spiral arms 
may be the direct consequence of 
cooperation among the Parker instability,  Jeans instability, and spiral arm 
potentials (e.g., \citealt{elm83}).

In \citet[hereafter Paper I]{kim01}, we presented studies of detailed
nonlinear evolution of self-gravitating modes in shearing, razor-thin 
galactic disks. We showed that formation of bound condensations
via the swing instability is subject to ``threshold'' behavior, with
models having sufficiently small values of the Toomre $Q$ parameter
(see eq.\ [\ref{Toomre_Q}] below) producing bound clouds, and models
with larger values of $Q$ remaining nonlinearly stable.  We found that the
threshold value of $Q$ is in the range $1.2-1.4$ (for magnetic field strengths
from zero up to thermal-equipartition values), consistent with
an apparent mean threshold $Q_{\rm th}=1.4$ for star formation evident 
from observational studies \citep{ken89,mar01}.
In Paper I, we also elucidated the role of MJI in forming bound 
condensations by showing that in low-shear regions where
the swing amplifier cannot operate, the mechanism that 
leads to growth of perturbations is quite different from that
in high-shear regions, and that the presence of magnetic fields is 
crucial to the instability.
In the absence of magnetic fields, specific vorticity is
conserved; contraction of vorticity-conserving
masses leads to increased Coriolis forces that exceed the
increase in self-gravity such that further contraction is prevented
(in the absence of a shearing background).
In magnetized systems, however, tension forces can transfer
vorticity from one region to another, such that condensation driven by
self-gravity can proceed.

Effects of spiral arm gravity on the growth of giant clouds were
explicitly addressed by \citet[hereafter Paper II]{kim02}
using two-dimensional models. Paper II showed that 
for a spiral arm inside corotation,
the enhanced surface density and shear gradients
in the arm region promote the growth of gaseous spurs. 
These spurs jut nearly
perpendicularly from the outer (downstream) side of the arm, and trail
in the same sense as the arm itself; this is consistent with
observed spur-like dust lanes in M51 \citep{sco01}.
We argued that the formation of spurs can be
understood as a natural consequence of the MJI within spiral arms, where
compression leads to a reduced (or negative) shear gradient that
allows these self-gravitating modes to grow over an extended time.
Paper II showed that bound clouds of typical mass 
$\sim 4\times 10^6 \,\Msun$ can form from spur fragmentation;
such condensations potentially represent
structures that would initially be observed as GMAs, and
subsequently evolve to become bright OB associations. 

Because of their restriction to razor-thin geometry, 
our previous two-dimensional models could not capture the potential 
consequences of the Parker instability on cloud formation. 
While previous work on the Parker instability has studied the weak-shear 
limit (see linear-theory analysis of \citealt{shu74}, \citealt{fog94,fog95})
and the case of solid-body rotation (see linear theory of
\citealt{zwe75} and simulations by \citealt{cho00} and \citealt{kimj01}),
and indeed shown that rotation and shear tend to be stabilizing,
no previous study focused primarily on Parker modes has incorporated shear 
at a realistic ``average'' galactic level.
Although the Parker instability has been proposed to prompt the formation
of molecular clouds (e.g., \citealt{mou_etal74,bli80}),
existing numerical simulations indicate that the Parker instability alone 
is unable to produce structures of high enough surface density
to represent GMCs 
\citep{bas97,kimj98_2,kimj00,san00,kimj01}.\footnote{In addition, 
a random component of 
galactic magnetic fields whose strength is almost comparable to the mean 
uniform component is shown to suppress the Parker instability
significantly \citep{kimj_ryu01}.}

To study the coupling of the Parker instability with self-gravitating modes, 
\citet{elm82a,elm82b} performed linear stability
analyses for combined Parker-Jeans modes and concluded that clouds of mass
$\sim 10^5-10^6\,\Msun$ form in the density regime for which self-gravity
significantly modifies the Parker instability; this regime of densities
is expected to be found in spiral arm regions only.
\citet{han92} included solid-body rotation in a linear-theory analysis
of the Parker-Jeans instability, and concluded that the transfer of angular 
momentum out of growing condensations may be a limiting factor. 
This transfer of angular momentum may be accomplished by magnetic stresses 
much as it occurs in the two-dimensional MJI.
A nonlinear simulation of the Parker-Jeans instability in a disk with
solid-body rotation was among the models performed by \citet{cho00},
who found that filamentary structures tend to form perpendicular to the
mean magnetic field. 

To our knowledge, there have previously been neither linear-theory
nor numerical studies that incorporate all of the aforementioned elements 
that can contribute to three-dimensional instabilities in galactic disks: 
buoyant (horizontal) magnetic fields, self-gravity, rotation, and {\it shear}.
In this paper, we investigate nonlinear evolution of galactic disks
allowing for all these fundamental physical agents. 
This work extends our previous studies (Paper I)
of structure formation via gravitational instability by including
the effects of the Parker instability.
Our primary objectives are to study how the Parker instability
develops in the presence of strong shear, and to compare the eventual
evolution of disks under the various instabilities separately or
in combination. We can thus assess the potential for
gravitationally-bound cloud formation from a variety of
dynamical mechanisms.
Since vertically-integrated, razor-thin disks are known to overestimate
self-gravity at the midplane (e.g., \citealt{too64}),
we will also study analytically and numerically how the stability of 
three-dimensional disks to in-plane self-gravitating modes changes as 
their thickness varies.

In our model, disks are local (see \S2.1), 
isothermal in both space and time, and 
vertically stratified. Initial magnetic fields are plane-parallel,
pointing in the azimuthal direction.
To model realistic conditions in disk galaxies,
we include both external gravity arising from stars (as a time-independent
potential) and gaseous self-gravity, 
which together counterbalance thermal and magnetic pressure
gradients in an unperturbed state.
We do not, however, consider in this paper other features 
such as stellar spiral arms and effects of cosmic rays that may 
enhance the magnetic buoyancy force.
We begin by solving the governing equations to
obtain initial static equilibrium density distributions,
and then explore their gravitational stability to axisymmetric
perturbations. We next select several parameter sets that 
represent ``average disk'' or ``spiral arm'' regions.
After applying small-amplitude perturbations to
the corresponding density profiles, we follow the nonlinear dynamical
development with three-dimensional direct numerical simulations.

The paper is organized as follows: In \S2, we describe the basic
MHD equations in the local ``shearing-sheet'' approximation, 
the computational methods we use, and our model
parameters. In \S3, we present initial magnetohydrostatic 
equilibria in the presence of both self-gravity and external gravity, 
and analytically examine
their axisymmetric stability by making simplifying assumptions. In \S4, 
we present the results of three-dimensional simulations 
for the Parker instability without self-gravity (\S4.1), generalized 
(Parker/swing) gravitational instability in shearing regions (\S4.2),
and MJI in nonshearing regions (\S4.3).
We summarize our results and discuss the implications of present
work for galactic cloud formation in \S5.

\section{Numerical Methods and Model Parameters}
\subsection{Basic Equations and Numerical Methods}

In this paper, we study nonlinear evolution of vertically stratified,
differentially rotating, self-gravitating galactic gas disks with 
threaded magnetic fields in the azimuthal direction.
We solve fully three-dimensional, compressible, ideal MHD equations 
in a local Cartesian reference frame whose center lies at a 
galactocentric radius $R_0$ and orbits the galaxy with a fixed angular 
velocity $\Omega\equiv\Omega(R_0)$. In this local frame, 
radial, azimuthal, and vertical coordinates are represented by
$x\equiv R-R_0$, $y\equiv R_0(\phi-\Omega t)$, and $z$, respectively,
and terms associated with the curvature effect are neglected 
\citep{gol65b,jul66}. The local-frame equilibrium background 
velocity relative to the center of the box at $x=y=z=0$ is then
reduced to
$\mathbf{v}_0 \equiv -q\Omega x \mathbf{\hat{y}}$, where
\begin{equation}\label{shear_q}
q\equiv -\frac{d\ln\Omega}{d\ln R}\bigg|_{R_0}, 
\end{equation}
is the local dimensionless shear rate.
In terms of $q$, the local epicyclic frequency $\kappa$ is
given by $\kappa^2 \equiv (4-2q)\Omega^2$.
The resulting ``shearing-sheet'' equations expanded in the local frame are
\begin{equation}\label{cont}
  \frac{\partial\rho}{\partial t} + \nabla\cdot(\rho \mathbf{v}) = 0,
\end{equation}
\begin{equation}\label{mome}
  \frac{\partial\mathbf{v}}{\partial t} +
        \mathbf{v}\cdot\nabla\mathbf{v} = -\frac{1}{\rho}\nabla P
       + \frac{1}{4\pi\rho}(\nabla\times\mathbf{B})\times\mathbf{B}
       + 2q\Omega^2 x \mathbf{\hat{x}} - 2\mathbf{\Omega}\times\mathbf{v}
       - \nabla \Phi_s + \mathbf{\gext},
\end{equation}
\begin{equation}\label{indu}
  \frac{\partial \mathbf{B}}{\partial t} =
       \nabla\times(\mathbf{v}\times\mathbf{B}),
\end{equation}
\begin{equation}\label{poss}
   \nabla^2\Phi_s = 4\pi G \rho,
\end{equation}
and 
\begin{equation}\label{eos}
   P = \cs^2\rho,
\end{equation}
(cf, \citealt{haw95, sto96}). Here, $\mathbf{v}$ is the
velocity in the local frame, 
$\rho$ is the mass density, $P$ is the thermal pressure, $\mathbf{B}$ 
is the magnetic field, $\Phi_s$ is the self-gravitational potential,
$\mathbf{\gext}$ is the external gravitational acceleration,
$G$ is the gravitational constant, and $\cs$ is the isothermal sound speed.
In equation (\ref{mome}), the third and fourth terms on the right hand side 
represent the tidal and Coriolis forces, respectively, which balance each 
other in the initial equilibrium with $\mathbf{v}=\mathbf{v}_0$. 
As expressed by equation (\ref{eos}), for simplicity we adopt an isothermal 
equation of state for all the simulations presented in this paper.

An initial vertical equilibrium is defined by the balance
between the total (thermal plus magnetic) pressure gradient and
the total (self plus external) gravitational force. 
For $\mathbf{\gext}$ representing the external gravity from stars, 
we adopt the simple form
\begin{equation}\label{extg}
   \mathbf{\gext} = - \frac{g_*}{H_*}\mathbf{z},
\end{equation}
where $H_*$ is the scale height of stars and $g_*$ is a free parameter.
Since most of gas is distributed 
within $|z| < H_*$ and the small-$|z|$ part of the realistic gravity 
deduced from stellar tracer populations can be described by a linear 
function of height (e.g., \citealt{bah86,kui89}), $\mathbf{\gext}$ given 
in equation (\ref{extg}) is a fairly good approximation in
studying dynamics of the gaseous component.
Although a linear gravity law is known to strengthen the Parker 
instability slightly compared to the case with a more realistic 
gravity law, the growth rate of the most unstable mode differs by 
only a factor of 1.2 \citep{kimj98_1}. 
As we shall show in \S 3, the total gravity that
drives the Parker instability in our models is similar to 
the external gravity adopted by other authors \citep{giz93,kimj98_1}.

The numerical solutions to the dynamical equations 
(\ref{cont})$-$(\ref{extg}) are obtained using a modified
version of the ZEUS code originally developed by \citet{sto92a, sto92b}.
ZEUS is a time-explicit, operator-split, finite-difference method
for solving the MHD equations on a staggered mesh, and has been
demonstrated to provide robust solutions to a wide variety of 
astrophysical problems. ZEUS employs ``constrained transport''
to guarantee that $\nabla\cdot\mathbf{B}=0$ within machine
precision, and the ``method of characteristics'' for accurate
propagation of Alfv\'enic disturbances \citep{eva88, haw_sto95}.
For less diffusive advection of hydrodynamic variables,
we apply a velocity decomposition method which treats
the contribution from the background shearing parts
as source terms, while solving the transport step
using only the perturbed parts.
In order to minimize the errors arising from
discontinuities in the flow characteristics across
the $x$-boundaries, the ghost zones adjoining the $x$-boundaries
are kept active (Paper I).
In the $\hat{x}$ and $\hat{y}$ directions, 
we implement shearing box boundary conditions in which the 
$x$-boundaries are shearing-periodic and
the $y$-boundaries are perfectly periodic
\citep{haw95}.
In the $\hat{z}$ direction, we adopt free (outflow) boundary conditions 
which set all the variables in the boundary zones
equal to the corresponding values in the adjacent active zones \citep{sto92a}.
These boundary conditions permit outgoing waves and thus physical 
quantities such
as mass, momentum, and magnetic flux can flow out of the simulation domain. 
Since density is very low near the vertical boundaries, however,
simulation results generally show that the total mass is conserved
within 1\%.

For computation of the self-gravitational potential, we employ a hybrid
technique that combines the Green function method
to compute the gravitational kernels 
\citep{miy87} and the Fourier transform method in shearing-sheet
coordinates \citep{gam01}.
Our high-resolution models have a $256\times128\times128$
grid for ($x$, $y$, $z$); we find that the Poisson solver typically takes 
about 48\% of the total cpu time. 
We have checked our implementation of the ZEUS code
on various test problems including two-dimensional
gravitational instability and propagation of gravito-MHD waves 
in three dimensions. 

\subsection{Model Parameters}

Our model disks are highly-idealized, smoothed versions of what is 
physically a much more complex, multiphase interstellar medium 
(e.g., \citealt{fie69,cox74,mck77,hei01}). We adopt an isothermal condition 
in both space and time, with an effective isothermal speed of sound
$\cs=7{\,\rm\,km\,s^{-1}}$, corresponding to a mean Galactic thermal 
pressure $P/k \sim 2000-4000{\,\rm K\,cm^{-3}}$ \citep{hei01}
and mean midplane density $n_{\rm H} \sim0.6\,{\rm cm^{-3}}$ \citep{dic90};
external galaxies have similar gaseous velocity dispersions.
One of the parameters in our simulations is the rotation rate $\Omega$.
When reporting dimensional results, we normalize to the 
solar neighborhood value 
$\Omega = 26 {\,\rm km\,s^{-1}\,kpc^{-1}}$ \citep{bin87}.
For a flat rotation curve, $q\approx 1$ and 
thus $\kappa\approx 2^{1/2}\Omega$; we allow for varying $q$ to 
represent both ``average'' regions and reduced-shear (spiral arm) regions.

Our initial disk models are vertically stratified, with
initial density distribution $\rho_0(z)$. 
The corresponding surface density is
given by $\Sigma_0\equiv \int_{-\infty}^\infty \rho_0(z)dz$.
To characterize $\Sigma_0$ in a dimensionless fashion, 
we use the Toomre $Q$ stability
parameter
\begin{equation}\label{Toomre_Q}
Q \equiv \frac{\kappa\cs}{\pi G \Sigma_0}.
\end{equation}
For an infinitesimally-thin disk, the critical
value for axisymmetric gravitational stability is $Q_c = 1$  
(\citealt{too64}; see also e.g., \citealt{bin87,shu92}).
We shall show in \S 3 that the finite thickness of disks reduces
$Q_c$ (cf.\ \citealt{gol65a}).
For local gas surface density of $\Sigma_0 = 13\,\Msun\,\pc^{-2}$ 
comparable to the solar neighborhood (cf, Paper I; \citealt{kui89}), 
$Q\approx 1.5$; we adopt this value of $Q$ for our 
``average disk'' fiducial value.

The initial magnetic field in our models points in the azimuthal 
($\mathbf{\hat{y}}$) direction and varies with height as 
$\mathbf{B}_0 = B_0(z) \mathbf{\hat{y}}$;
its magnitude is parameterized by 
\begin{equation}\label{beta}
\beta \equiv \frac{\cs^2}{\vA^2}, 
\end{equation}
where the Alfv\'en speed is $\vA\equiv B_0/\sqrt{4\pi\rho_0}$. 
As an initial condition,
we fix $\beta=1$ spatially for all the MHD models in this paper,
so that initially the Alfv\'en speed is the same as the isothermal
sound speed everywhere. This implies $B_0(z) \propto \sqrt{\rho_0(z)}$.

We explore the nonlinear evolution of disks under both average disk
and other galactic conditions by changing $\Sigma_0$ and $q$. 
In order to model a portion of a gas disk inside a spiral arm,
we use $\Sigma_0 = 65 \,\Msun\,\pc^{-2}$,
which is five times larger than the average disk value.
Since $\kappa\propto \Sigma^{1/2}$ and $Q\propto \Sigma^{-1/2}$ 
for isothermal gas from the constraint of potential vorticity conservation 
(\citealt{bal85}; Paper II), this higher surface density
corresponds to $Q \approx 0.7$, which we adopt as a fiducial ``spiral arm''
value. We allow the shear parameter
$q$ to vary from model to model, but fix it spatially in a given model.
Note that this arbitrary implementation of background shear is not consistent 
with realistic shear that varies rapidly across the spiral arm (Paper II).
Nevertheless, our high surface density models allow us to focus on the 
effects of enhanced self-gravity inside the spiral arm without
the additional complications of in-plane variations of the background.

We parameterize the magnitude of the external gravity by 
\begin{equation}\label{s0}
 s_0 \equiv \frac{4\pi G\rho_\infty H_*}{g_*},
\end{equation}
which measures the relative strength of self-(gaseous) gravity and external
(stellar) gravity in the equilibrium state (cf, \citealt{elm82a}). Here, 
$\rho_\infty$ denotes the midplane density in
the absence of both magnetic field and external gravity
(i.e., $\beta=\infty$ and $s_0=\infty$; see \S 3).
If the stellar distribution approximates a vertical
self-gravitating equilibrium with vertical velocity dispersion $\sigma_{*,z}$,
one can rewrite equation (\ref{s0}) as
\begin{equation}\label{s0_new}
s_0 = \left(\frac{\sigma_{*,z}\Sigma_0}{\cs\Sigma_*}\right)^2, 
\end{equation}
where $\Sigma_*$ denotes the stellar surface density.\footnote{For an 
application to a central-mass dominated Keplerian disk, 
$\gext =-\Omega^2 z$. Then, $s_0\equiv 4\pi G\rho_\infty/\Omega^2 = 2 Q^{-2}$.}
Notice that $s_0$ is now expressed in terms of observable quantities. 
Since $\Sigma_*\approx35 \,\Msun\,\pc^{-2}$
and $\sigma_{*,z}\approx20 {\,\rm\,km\,s^{-1}}$ (e.g., \citealt{kui89,
hol00}), $s_0\approx 1$ is suitable for ``average
disk'' conditions, while $s_0\approx 25$ would describe a 
spiral arm region (taking the stellar spiral perturbation as weak).

Our simulation domain is a rectangular parallelepiped of size 
$L_x\times L_y\times L_z$, with a vertical range of $|z|\leq L_z/2=4H$, 
where $H$ is the gaseous scale height (see eq.\ [\ref{scal}] below). 
The horizontal dimensions of the box are $L_x=L_y=25H$ or $17H$
for $s_0=1$ or $s_0=25$ models, respectively. This size corresponds
to the most unstable wavelength of the axisymmetric gravitational
instability in each model with $\beta=1$ (see \S3.2).
Conversion of $H$ into a physical value requires solving
equations governing magnetohydrostatic equilibria, which will be 
presented in the next section. Using $\cs/\Omega = 270\,{\rm pc}\,
(\cs/7\,{\rm km\,s^{-1}})(\Omega/26 \,{\rm km\,s^{-1}\,kpc^{-1}})^{-1}$
and borrowing the results from Figure \ref{H_s0}, 
we obtain for high surface-density (spiral arm) regions ($s_0=25$),
$H\approx50$ pc for $\beta=\infty$ and $H\approx80$ pc for $\beta=1$ models.
Under average disk conditions ($s_0=1$),
$H\approx160$ pc for $\beta=\infty$ and $H\approx210$ pc for $\beta=1$ cases.
The former value for an unmagnetized model is consistent with 
the usual inner-galaxy scale height of $\sim$130-170 pc 
derived from the hydrostatic force balance between external gravity
and turbulent pressure in the average disk condition 
(see e.g., \citealt{bou90,loc91,kimj00}). 
On the other hand, the latter value for a $\beta=1$ model is similar to the 
direct estimate $\sim$220 pc for the \ion{H}{1} scale height 
obtained by tangent point observations near the solar circle \citep{mal95},
beyond which the \ion{H}{1} disk is known to flare dramatically
\citep{kul82}.

The larger initial scale height of magnetized models reflect the
potential importance of magnetic pressure in supporting a galactic disk
against gravity. If magnetic support is significant but is 
neglected in observationally-based determinations of the scale height
from surface density and velocity dispersion, then $H$ could be 
underestimated. We note, however, that the dynamical rearrangement
of magnetic flux in our simulations always results in a reduction of
the scale height over time, so that the initial values of $H$ in
our $\beta=1$ models is an upper limit for a real disk of
high magnetization.

The total mass 
$M_{\rm tot} \equiv \Sigma_0L_xL_y$ contained in the simulation box
corresponds to $\sim3.6\times10^8\Msun (H/210 {\rm pc})^3$ 
for $s_0=1,\; \beta=1$ models, 
$\sim1.2\times10^8\Msun (H/80 {\rm pc})^3$ 
for $s_0=25,\; \beta=1$ models, and
$\sim0.5\times10^8\Msun (H/50 {\rm pc})^3$ 
for $s_0=25,\; \beta=\infty$ models.

For all the simulations reported in this paper, we apply 
initial perturbations only to the density. We first 
generate a Gaussian random variable $\delta$ using the usual
Box-Muller algorithm (cf, \citealt{pre92}) and then construct
its power spectrum in the Fourier space such that 
$\langle|\delta_k|^2\rangle \propto k^{-11/3}$ for
$1 \leq kL_x/2\pi \leq 32$ and zero for $32 < kL_x/2\pi$.
We normalize by fixing the standard deviation of $\delta$ in real 
space to be 1\%.
The final form of density perturbations is given by the product 
of $\delta$ and the background density profile $\rho_0$.

Finally, we remark on a few dynamical timescales of note. 
The rotation time is defined by
$t_{\rm rot}\equiv 1/\Omega=3.8\times 10^7\,
{\rm yrs}\,(\Omega / 26\,{\rm km\,s^{-1}\,kpc^{-1}})^{-1}$.
The shearing time is $\tsh \equiv 1/q\Omega$, which
is equal to $t_{\rm rot}$ for a flat rotation curve ($q=1$).
Related to $\tsh$ is the swing amplification time scale, 
which is approximately $\tswing\sim (4-5)\tsh$ 
(cf, \citealt{too81}; Paper I).
The orbital period is  
$\torb \equiv 2\pi/\Omega = 2.4\times 10^8\,{\rm yrs}\,
(\Omega / 26 \,{\rm km\,s^{-1}\,kpc^{-1}})^{-1}$,
which is the time unit we adopt for our presentation.
The typical growth time of the Parker instability
is given by the Alfv\'en crossing time over the scale height
\begin{equation}\label{tpar}
\tParker \equiv \frac{H}{\vA} = 3.0\times 10^7\;{\rm yrs}\,
\left(\frac{H}{210\;{\rm pc}}\right)
\left(\frac{\vA}{7\,{\rm km\,s^{-1}}}\right)^{-1},
\end{equation}
while a characteristic gravitational contraction time is 
\begin{equation}\label{tgra}
\tgrav \equiv \frac{\cs}{G\Sigma_0}=
   1.2 \times 10^8\,{\rm yrs}
    \left(\frac{\cs}{7 {\rm\,km\,s^{-1}}}\right)
   \left(\frac{\Sigma_0}{13\Msun\pc^{-2}}\right)^{-1},
\end{equation}
(Paper I). Note that the Parker instability involves vertical
motions of gas, while fluid motions in the gravitational 
instability are mostly parallel to the galactic plane 
at least in the linear regime.

\section{Initial Equilibria and Axisymmetric Stability}
\subsection{Initial Magnetohydrostatic Equilibria}

In the absence of perturbations, our isothermal disk is 
in vertical magnetohydrostatic equilibrium with a density
profile $\rho_0(z)$. Thermal and magnetic
pressure forces balance against the combination of gaseous self-gravity and 
external gravity. Stronger gravity leads to a thinner disk.
In this section, we study the change of disk thickness as $s_0$ varies.

For simplicity, let us define $\dmid$ be the midplane value
of $\rho_0$ and 
\begin{equation}\label{scal}
H\equiv \frac{1}{\dmid}\int_0^{\infty} \rho_0(z) dz,
\end{equation}
be the gaseous scale height for general cases,
while $\rho_\infty$ and $H_\infty$ are the counterparts for 
purely self-gravitating, unmagnetized disks
(i.e., $s_0=\infty$ and $\beta=\infty$).
From the definitions, it follows for fixed $\Sigma_0$ 
that $H=\Sigma_0/(2\dmid) = H_\infty\rho_\infty/\dmid$
and using the \citet{spi42} isothermal disk solution 
\begin{equation}\label{H_inf}
H_\infty \equiv \frac{\cs}{(2\pi G \rho_\infty)^{1/2}}
= \frac{\cs^2}{\pi G \Sigma_0}
 = 280 \,{\rm pc} 
 \left(\frac{\cs}{7 {\rm\,km\,s^{-1}}}\right)^2
   \left(\frac{\Sigma_0}{13\Msun\pc^{-2}}\right)^{-1}.
\end{equation}
Arranging equations (\ref{mome}) and (\ref{poss}) -- (\ref{extg}) for
a static equilibrium in the vertical direction,
we are left with 
\begin{equation}\label{vert}
  \frac{1}{2}\left(1 + \frac{1}{2\beta}\right)
  \frac{d^2\ln \tilde{\rho}}{d\zeta^2} 
 = - \frac{H_\infty}{H}\tilde{\rho}
   - \frac{1}{s_0},
\end{equation}
where the dimensionless height $\zeta\equiv z/H_\infty$ and
the dimensionless density $\tilde{\rho}(z) \equiv \rho_0(z)/\dmid$.
Equation (\ref{vert}) is to be integrated subject to the
constraint (\ref{scal}).

In the two limiting cases of $s_0$, equations (\ref{scal}) and 
(\ref{vert}) yield analytic solutions.
For $s_0 \gg 1$ corresponding to a strongly self-gravitating disk
and negligible external gravity, we obtain 
$\tilde{\rho} = {\rm sech}^2 (z/H)$ along with 
$H/H_\infty = 1 + 1/2\beta$ (e.g., \citealt{spi42,elm78}), while for
$s_0 \ll 1$  corresponding to weak self-gravity compared to
external gravity,
$\tilde{\rho}= \exp(-\pi z^2/4H^2)$ along
with $H/H_\infty = (\pi s_0)^{1/2}(1+1/2\beta)^{1/2}/2$.
Note that the latter Gaussian function is the usual solution for
density in an equilibrium under linear gravity, with
$4H^2/\pi = 2\cs^2H_*/g_*$.
      
We solve equations (\ref{scal}) and (\ref{vert}) iteratively:
for general $s_0$, we arbitrarily choose an initial $H/H_\infty$ and
numerically integrate equation (\ref{vert}) to find 
$\tilde{\rho}(\zeta)$. If the calculated $\tilde{\rho}(\zeta)$
does not satisfy the constraint (\ref{scal}),
we modify $H/H_\infty$ and iterate
until convergence is attained. The resulting $H/H_\infty$
as a function of $s_0$ is shown in Figure \ref{H_s0}
where the solid line corresponds to the
unmagnetized case ($\beta=\infty$), while dotted and dashed lines are 
for sub-equipartition ($\beta=10$) and equipartition ($\beta=1$)
magnetic field cases, respectively. 
To compare $H$ with a length scale independent of the gas disk's mass,
Figure \ref{H_s0} also plots 
$H\Omega/\cs = (H/H_\infty)(Q\Omega/\kappa)$ as a function of $s_0$.
One may think of increasing $s_0$ 
as either decreasing the stellar content ($\Sigma_*$) while holding 
the gas content ($\Sigma_0$) fixed, or 
increasing $\Sigma_0$ for a given value of $\Sigma_*$. Since
$H_\infty$ is dependent on $\Sigma_0$, the first meaning
of $s_0$ is helpful in interpreting the $H/H_\infty$ curves: 
moving toward larger (smaller) $s_0$,
a gas disk thickens (thins) as the stellar content relative to gas 
decreases (increases).
The magnetic pressure support has a larger impact at higher $s_0$.
On the other hand,
since $\Omega/\cs$ is independent of $s_0$ and 
assumed to be constant at a given radius in a galaxy, the second meaning
of $s_0$ helps interpret the $H\Omega/\cs$ curves:
as gas moves from interarm regions into spiral arms where density
and $s_0$ are higher, the gas layer 
becomes thinner in response to its own enhanced self-gravity. 
For instance, $H(s_0=1)/H(s_0=25)\approx 3$ when $\beta=\infty$.
Note that to take into account the effects
of compressed magnetic fields, $\beta$ should
change correspondingly in spiral arms.

As mentioned before, the effective driving force, $\geff$, of 
the Parker instability in our models is the sum of 
the external gravity and the self-gravity of a disk,
\begin{equation}\label{geff}
\geff \equiv \gext - \frac{d\Phi_s}{dz}.
\end{equation}
In Figure \ref{ext_grav} we plot $\geff$ as solid lines 
together with contributions of gaseous self-gravity (dotted lines)
and external stellar gravity (short-dashed lines). 
For average disk conditions ($Q\approx 1.5$ and $s_0\approx 1$), 
for $|z|/H\simlt 1$, external gravity is as important as 
self-gravity in establishing an equilibrium (Fig.\ \ref{ext_grav}$a$).
For comparison, we also plot the external gravity models
adopted by \citet{giz93} and \citet{kimj98_1}, which
are in good agreement with our total average disk gravity model.
On the other hand, strongly-compressed
spiral arm regions are entirely 
dominated by self-gravity (Fig.\ \ref{ext_grav}$b$).

\subsection{Axisymmetric Stability}

Dynamics of galactic gaseous disks have often 
been studied by using height-integrated equations,
accurate when the vertical height of the disk is much smaller than
the physical scale of interest (e.g, \citealt{bin87,shu92}).
One of the drawbacks of this thin-disk approximation is
that it overestimates self-gravity for horizontal scales approaching
the disk thickness (e.g., \citealt{too64}). 
As mentioned earlier, the thin-disk approximation leads to
$Q_c=1$ for axisymmetric gravitational instability.
\citet{gol65a} studied the stability of uniformly rotating disks of
finite thickness, and
showed that $Q_c=0.676$ for a {\it purely} self-gravitating 
($s_0=\infty$), isothermal disk (see also \citealt{gam01}). 
In this section, we explore the axisymmetric stability of 
isothermal disks under both self-gravity and external gravity.

For analytic simplicity, we make an approximation that fluid motions are all 
lateral, confined to the $x$--$y$ plane. 
This ignores the small vertical oscillations of the disk as well as 
the potential effects of the Parker instability 
(which would be nonaxisymmetric anyway).
We perform a standard linear stability analysis for axisymmetric modes,
assuming perturbations are a linear superposition of Fourier
modes, $\exp(i\omega t-ikx)$, of frequency $\omega$ and wavenumber $k$.
For the gravitational force due to a perturbed density distribution
$\rho_1=\rho_1(z)e^{-ikx}$
with $k\neq0$, the Poisson equation (\ref{poss}) yields an approximate 
solution $d\Phi_1/dx = 4\pi i{\,\rm sgn}(k)G\rho_1(0) He^{-ikx}/(1 + |k|H)$
at the disk midplane. This expression for the 
perturbed force is exact for an exponential 
density distribution $\rho_1(z)\propto e^{-z/H}$ (e.g., \citealt{elm87}),
and asymptotically approaches the exact 
solutions for arbitrary density profiles when $|k|H\rightarrow0$ or 
$\infty$. For the vertical density profiles that we found in the previous 
section, the largest fractional difference between the true and approximate
values for the perturbed radial gravity is only $\sim 15\%$, which occurs 
at $|k|H\sim1$. 
The procedure for driving the dispersion relation is quite straightforward 
and we write down only the result,
\begin{equation}\label{disp}
\omega^2=\kappa^2+(\cs^2+\vA^2)k^2-\left(\frac{|k|H}{1+|k|H}\right)
4\pi G\dmid. 
\end{equation}
For short-wavelength perturbations $kH \gg1$, equation (\ref{disp}) recovers 
the dispersion relation
of waves in a rotating medium with uniform density $\dmid$ in 
three dimensions \citep{cha61}. For long-wavelength perturbations 
$kH \ll 1$, equation (\ref{disp}) reduces to the dispersion relation in a 
razor-thin disk with surface density $\Sigma_0=2\dmid H$.

Using the $H-s_0$ relationship that we found in the previous subsection, 
we compute $\kmax$ which minimizes the right hand side of equation 
(\ref{disp}) for given $s_0$ and $\Sigma_0$ (and thus maximizes the
growth rate). The results are shown in Figure \ref{Qc_s0}$a$.
The most unstable wavelength for axisymmetric perturbations is typically
$\lambda_{\rm max} = 2\pi/\kmax \sim 20H$ for average disk conditions
(i.e., $s_0=1$) and decreases slightly as gaseous self-gravity
becomes more important (larger $s_0$). 
For fixed gas content, the $\kmax H_\infty$ curves show that
relatively weaker external gravity (larger $s_0$) requires a larger spatial
scale for perturbations to become unstable. Note that 
$2/H_\infty=2\pi G\Sigma_0/\cs^2$ is equal to the Jeans wavenumber of 
a razor-thin disk (cf, Paper I).

We calculate $Q_c = \kappa (2\pi G\rho_\infty)^{-1/2}$ corresponding to 
marginal stability modes ($\omega=0$) from equation (\ref{disp}) and \S3.1, 
and plot the results in Figure \ref{Qc_s0}$b$. Our simplified
analytic approach gives $Q_c=0.647$ for unmagnetized disks with
no external gravity ($s_0\rightarrow\infty$), which is close to
$Q_c=0.676$ of \citet{gol65a}'s rigorous analysis.
Critical values $Q_c$ for other $s_0$ and $\beta$ are
shown in Figure \ref{Qc_s0}$b$.
We also perform axisymmetric ($x$--$z$) simulations using our numerical code 
to find $Q_c$ for realistic density distributions obtained in \S3.1.
The simulation results are displayed in  Figure \ref{Qc_s0}$b$ 
as filled circles ($\beta=\infty$), triangles ($\beta=10$), and 
squares ($\beta=1$),
which are in excellent agreement with the results of the analytic approach.
For average disk conditions ($s_0=1$),
$Q_c$ is found to be $\sim$0.75, 0.72, and 0.57 for $\beta=\infty$,
10, and 1 cases, respectively. These represent
approximately a 25\% reduction relative to the corresponding $Q_c$ 
values of razor-thin disks. 
This suggests that the average disk with $Q\approx 1.5$
is quite stable to axisymmetric gravitational instability. 

We do not attempt in this paper to pinpoint threshold $Q_{\rm th}$ values 
for gravitational runaways due to (nonaxisymmetric) swing amplification 
in finite-thickness disks, as we did in Paper I for razor-thin disks; there
we found $Q_{\rm th}=$1.3, 1.4, and 1.2 for $\beta=\infty$, 10, and 1 cases,
respectively. If we naively assume that $Q_c$ is lower by $\sim25\%$ 
in nonaxisymmetric cases as well, then thick disks would have
$Q_{\rm th}\sim 0.9-1$; our results from \S4 would be consistent with
such thresholds. The reduction of self-gravity 
in finite-thickness disks also implies that stable quasi-axisymmetric spiral
shock solutions can exist at a stronger spiral arm potential than 
predicted in the razor-thin disks (cf, Paper II). 

\section{Three-Dimensional Simulations}

To investigate local three-dimensional evolution of galactic gas disks under
various conditions, we have performed a total of 10 numerical
simulations. Table \ref{tbl-1} lists the important parameters 
for each model. Column (1) labels each run. Column (2) gives the
Toomre parameter $Q$ (see eq.\ [\ref{Toomre_Q}]). Note that $Q<1$
in a thick disk does not necessarily mean axisymmetric  gravitational
instability (see \S3 and Fig.\ \ref{Qc_s0}). 
The relative strength of self-gravity compared to external
gravity is given in terms of $s_0$ (see eq.\ [\ref{s0}]) in column (3);
$s_0=25$ models strong spiral arm regions, while $s_0=1$ 
characterizes average disk conditions. 
Column (4) lists the plasma parameter $\beta$
characterizing the magnetic field strength (see eq.\ [\ref{beta}]).
The shear parameter $q$ (see eq.\ [\ref{shear_q}]) is given in column (5).
Column (6) indicates whether 
self-gravity is implemented in the numerical evolution. 
Notice that even for the cases with ``no'' self-gravity for the evolution, 
the vertical total gravity profile from the ``initial'' equilibrium 
is included in order to keep the unperturbed equilibrium density 
profile intact. Columns (7) and (8)
are our numerical resolution and the physical size of the
simulation box in units of $H$. 
The relationship of $H$, $Q$, and $s_0$ to physical values is
given in \S2.2. Finally, column (9)
describes the relevant physics in each model. 

\subsection{The Parker Instability Without Self-Gravity}

In this section, we describe the nonlinear evolution of the Parker
instability\footnote{Precisely speaking, the Parker instability that has
the wavenumber vector $\mathbf{k}$ parallel to the initial magnetic
field $\mathbf{B_0}$ is an {\it undular} mode of a more general 
class of magnetic buoyancy instabilities.
The other branch of magnetic buoyancy instability is 
an {\it interchange} mode whose wavenumber vector is perpendicular to 
$\mathbf{B_0}$ (e.g., \citealt{hug87}).
Since the interchange mode requires $\gamma < 1-1/(2\beta)$ to
be unstable, where the $z$-independent plasma
$\beta$ is defined by eq.\ (\ref{beta}) 
and $\gamma$ is an adiabatic index (cf, \citealt{kimj98_2}), 
our isothermal MHD models are all stable to
the interchange mode.}
in the absence of self-gravity. These non-self-gravitating 
models demonstrate the effects of rotation and shear on the Parker 
instability, and will later be compared with the results of 
self-gravitating models. 
From the outset, we should mention that magnetorotational instability
(MRI) in shearing disks with the present parameters is not important 
compared to the Parker instability. 
It is well known that MRI of toroidal magnetic
fields grows much slower than MRI of poloidal fields 
(cf, \citealt{bal98,kim00}). Even in a stratified disk with
$q=1.5$ and $\beta=12.5$, toroidal MRI takes about ten 
orbital times to fully grow \citep{sto96}.
Since the present models have 
weaker shear and stronger magnetic fields and are
evolved only up to $\sim(4-5)\torb$, effects of MRI are negligible.

\subsubsection{The Parker Instability in Uniformly Rotating Disks}

We begin by considering the Parker instability under rigid-body 
rotation (model A). Figure \ref{s25_evol}$a$ shows the evolutionary
history of the density maximum as a solid line. 
A three-dimensional visualization of the perturbed density 
with representative magnetic field lines is shown in the left panel of 
Figure \ref{parker3d}. Snapshots of total surface 
density as well as $x$--$y$ density slices are shown in
Figure \ref{col1}.

Initially, density perturbations which are a superposition of 
numerous modes with different $\mathbf{k}$ adjust the disk by launching
MHD waves. During this relaxation process, fluid motions are very active at 
high-$|z|$ regions where the density is low. Once the system finds
the most unstable Parker mode, over-dense regions slide toward the
midplane along the undulating field lines, 
while under-dense regions move upward 
at the expense of gravitational potential energy (cf, \citealt{hug87}), 
causing the field lines to bend further. 
Using the evolution of vertical velocity dispersion, we find 
the growth rate of the most unstable mode is $\sim 0.4\cs/H$ in model A.
After the maximum
density $\dmax=1.75\dmid$ is reached at $t/\torb\approx 1.8$, 
the Parker instability stops growing and saturates; 
the corresponding maximum surface density is $\Sigma_{\rm max}=1.35\Sigma_0$.
The primary reason for this is of course that the Parker instability
is not a runaway process, but is eventually stabilized by tension
forces from bent magnetic field lines (cf, \citealt{mou74}).

As the left frame of Figure \ref{parker3d} and Figure \ref{col1}$a$ show,
the dominant growing mode turns out to be {\it antisymmetric} across the
midplane, and {\it trailing} with $\kx=9(2\pi/L_x)=3.3/H$ and 
$\ky=2(2\pi/L_y)=0.74/H$; the corresponding wavelengths are $\lambda_x=2H$
and $\lambda_y=8.5H$. The antisymmetric (in density) property of the dominant
Parker mode in our model disks is consistent with 
the results of \citet{san00}, who showed that 
numerical simulations with random initial perturbations 
generally exhibit antisymmetric modes,
even though the linear growth rates for symmetric and antisymmetric modes 
of the Parker instability are almost indistinguishable in 
galactic environments \citep{kimj00}.\footnote{\citet{hor88} showed 
that the growth rates of antisymmetric and symmetric modes depend on 
their $\epsilon$ parameter (see eq.\ [10] of \citealt{hor88}). 
In galactic conditions, however,
$\epsilon \approx 10^3\, (R_0/10\,\kpc)^2 
(\Omega/26 {\,\rm km\,s^{-1}\,kpc^{-1}})^2 
(\cs/7 {\,\rm km\,s^{-1}})^2$, so that
their Fig.\ 5 reads that the maximum growth rates of antisymmetric
modes are almost the same as those of symmetric modes.} 

The dominant mode in model A is trailing ($\kx/\ky>0$), as well.
Linear dispersion relations for the Parker instability with
rotation are
independent of the sign of $\kx/\ky$ (cf, \citealt{shu74}),
and we have found that for low-resolution simulations seeded with 
either leading or trailing modes, growth rates are similar.
However, for the random perturbation form we adopt, trailing modes grow
sooner because of a shorter readjustment phase, for the rotating cases.
Additional tests for the nonrotating case show that 
the growth of trailing and leading modes is indistinguishable.

Three-dimensional Parker instability 
attains its maximum growth rate at $\kx\rightarrow\infty$, because 
larger $\kx$ provides more space for a raised section of a flux tube 
to expand in the radial direction \citep{par67}. As pointed out
by \citet{shu74}, however, the growth rate of $\kx=\infty$ mode 
is not very different from that of the $\kx=0$ modes and the growth
rate is essentially constant for $\kx H \simgt 5$. 
Since our high resolution simulations can resolve modes up to 
$\kx H \sim 12$, the dominance of the $\kx H=3.3$ modes in model A
is physical rather than numerical. 
The dominance of this particular mode over potential rivals may in part
be attributed to the shape of initial perturbation spectrum which has larger 
power at smaller $\kx$, while larger-$\kx$ modes, if resolved, grow faster;
a compromise may then select intermediate-$\kx$ modes that have
large enough initial power and fast enough growth rates.
It is also possible that nonlinear effects help to select this
mode, perhaps preferring $\lambda_x$ close to the full disk thickness.
We note that although uniform rotation tends to stabilize the 
Parker instability since the Coriolis force couples azimuthal 
and radial motions,
its effect is not significant if $\kx/\ky \gg 1$ 
(cf, \citealt{shu74,zwe75,fog94}).

In addition to helping select trailing modes for initial growth, 
the effects of rotation are evident in other details of the model 
evolution. As material falls toward the midplane along wavy 
magnetic field lines, it also gains speed in the $\mathbf{\hat{y}}$
direction, and its path is consequently
deflected in the $\mathbf{\hat{x}}$ direction by the Coriolis force. 
Gas moving toward a magnetic
valley with positive (negative) $v_y$ is redirected radially outward 
(inward). The net result is that over-dense regions at valleys tend to 
rotate in the counterclockwise direction, which in turn twists magnetic
field lines in the valleys (cf, \citealt{kimj01}). 
At the point when the instability begins to saturate,
gas at the valleys is slightly overcompressed and begins to
reexpand vertically. At the same time, the fields lines that previously
became twisted due to the Coriolis forces now exert back torques, 
inducing gas motions in the 
clockwise direction. The meridional flows at the disk midplane, as
displayed in Figure \ref{col1}$c$, 
show this sort of behavior. The flows are generally
subsonic near the midplane, while high altitude regions ($|z|/H > 2.5$)
have supersonic velocities. 

Evolution of the system following saturation is rather complicated,
involving nonlinear interactions of numerous MHD waves.
Overpressured, dense regions shown in 
Figure \ref{col1}$b$ soon expand in the horizontal direction. 
Expanding material interacts with gas that moves along the 
vertical direction. These active, but still mostly subsonic, flow
motions eventually homogenize the density,
resulting in a smooth surface density distribution
(Fig.\ \ref{col1}$d$). The maximum surface density after 3 orbits
is only 1.13 times the initial value.

The small-scale chaotic flow associated with nonlinear development
of large-$|\mathbf{k}|$ modes is also likely to cause
reconnection of the curved and twisted
magnetic field lines, preferentially at low-$|z|$ regions 
(cf, \citealt{kimj98_2,kimj01}). 
This in turn reduces the field strength near the 
midplane, which was already decreased during the
process of (midplane-crossing) Parker instability. 

We define the differential mass 
\begin{equation}\label{dmass}
\frac{dM(z)}{dz} \equiv \int\int \rho dx dy,
\end{equation}
and the differential magnetic flux averaged along 
the $y$-direction
\begin{equation}\label{dflux}
\frac{d\Phi_B(z)}{dz} \equiv \frac{1}{L_y}\int\int B_y dx dy,
\end{equation}
and plot $dM/dz$, $d\Phi_B/dz$, and $dM/d\Phi_B$ as functions of $z$ 
in Figure \ref{mass2flux0} for five time epochs.
For $t/\torb\simlt1$, both profiles remain unchanged. As the
Parker instability develops considerably ($1\simlt t/\torb\simlt2$), 
field lines float upwards, supplying magnetic flux at high 
altitude regions; magnetic fields may escape the simulation box
because of the outflow $z$-boundary conditions we adopted. 
Magnetic fields rise
up and gas slips down, enhancing the mass-to-flux ratio further
at the midplane (Fig.\ \ref{mass2flux0}$c$), and evening out
the $z$-distribution of magnetic flux (Fig.\ \ref{mass2flux0}$b$),
but mass enhancement near the midplane due to the Parker instability is 
generally quite moderate (Fig.\ \ref{mass2flux0}$a$).
At the end of the simulation, the overall density profile in model A
is very close to another hydrostatic equilibrium where
the initial effective external force ($\geff$; see eq.\ [\ref{geff}]) 
counterbalances the thermal pressure gradient alone. 
As Figure \ref{mass2flux0}$b$ suggests, 
the mean magnetic field (which is predominantly azimuthal) is
nearly constant with height, and therefore does
not contribute much in the final force balance. 
At $t/\torb=4$, the density-weighted velocity dispersions are calculated to be 
$\sigma_x \equiv \langle \rho v_x^2\rangle^{1/2}/\langle\rho\rangle^{1/2}
\approx 2.5\times 10^{-2}\cs$, 
$\sigma_y \equiv \langle \rho v_y^2\rangle^{1/2}/\langle\rho\rangle^{1/2}
\approx 7.3\times 10^{-2}\cs$, and 
$\sigma_z \equiv \langle \rho v_z^2\rangle^{1/2}/\langle\rho\rangle^{1/2}
\approx 2.4\times 10^{-2}\cs$, where
$\langle\;\rangle$ denotes a volume average.
Given the low level of the velocity dispersion, the Parker
instability is unlikely a significant source of interstellar turbulence in 
galactic disks.

\subsubsection{Effects of Differential Rotation} 

One of the crucial impacts differential rotation makes on the evolution 
of the Parker instability is that it 
changes the power spectrum of initial perturbations. In a shearing disk,
the wavefront of any disturbance evolves in response to the kinematics of 
the background flow; for uniform shear with $q\equiv - d\ln\Omega/d\ln R$,
the radial wavenumber of any pattern linearly increases with time as
$\kx = \kx(0) + q\Omega\ky t$. 
In our chosen initial density perturbation spectrum, 
the mode with $\kx(0)=0$ and $\ky= 1(2\pi/L_x)=0.37/H$ has much larger
power than higher-$\kx$ modes at the same $\ky$. 
The higher-$\kx$ modes, however, have higher growth rates than the 
$\kx=0$ modes. As a consequence of the initial power distribution,
non-shearing model A thus has to spend about one orbital time 
before the (large-$\kx$) Parker modes grow enough to produce
significant density fluctuation (Figure \ref{s25_evol}$a$). 
On the other hand, the background shear in model B
can transform the initial $\kx(0)=0$ mode, without changing its power, into a 
higher-$\kx$ disturbance -- for which the Parker instability is more
efficient. Therefore, in Figure \ref{s25_evol}$a$ the initial growth of 
perturbations is more evident in model B than in model A.
Structural analysis shows that this initially growing mode in model B 
is a {\it $z$-symmetric} (in density) mode with $\ky H= 0.37$ 
($\lambda_y=17H$) and confined to $|z|/H<1$. 
In contrast, the most unstable mode in model A is a {\it $z$-antisymmetric}
mode with $\ky H= 0.74$. 
The growth rate of the $\ky H=0.37$ mode in model B is found to be 
$\sim0.2\cs/H$, which is just half of the growth rate of model A.

While the initial evolution of shearing model B is governed by the 
growth of $z$-symmetric modes near the midplane,  
$z$-antisymmetric perturbations that have lower initial power 
(since $k_z$ is larger) also grow in high altitude regions at double the rate.
As they grow, over-dense regions fall down toward the midplane
and begin to interact with the symmetric modes. 
The right panel of Figure \ref{parker3d} displays perturbed density 
and magnetic field lines in model B at $t=10\Omega^{-1}$.
The condensations evident at the midplane are associated with the
symmetric mode with $\kx=10(2\pi/L_x)$ ($\lambda_x=2H$). 
Below or above these are ``tadpole-shaped'' condensations associated with 
the antisymmetric mode having $\kx=20(2\pi/L_x)$ ($\lambda_x=H$).
Magnetic field lines from model B appear straighter than from model A, 
suggesting that the (longer-wavelength) symmetric modes partially prevent
the (shorter-wavelength) antisymmetric modes from crossing the midplane.
As antisymmetric modes grow dominant later on ($t/\torb > 1.7$), 
field lines become progressively more curved, but 
never to the extent seen in model A.

For shearing model B, antisymmetric modes culminate their growth 
at $t/\torb\approx 2.1$. After model saturation,
the system experiences vigorous interactions between symmetric
and antisymmetric modes. An added complication
is the background shear, which stretches out any condensation that forms.
We plot snapshots of surface density $\Sigma/\Sigma_0$ in Figure \ref{q1_col},
which shows that at $t/\torb=2$ the system is dominated by high-$\kx$ modes,
while surface density at $t/\torb=3$ is rather uniformly distributed.
To some extent, the smooth surface density is attributable to
our limited numerical resolution. However, 
even if the true spatial scale at which dissipative processes smooth out
strongly phase-mixed structures is small compared to our numerical
resolution, there are evidently no nonlinear feedback processes to drive
late-time growth of large-scale perturbations.
Overall, the late time character of the system is 
a new, large-scale vertical equilibrium containing small-scale, 
fluctuating density and velocity fields. 
The density-weighted dispersions of the perturbed velocity in model B at 
time $t/\torb=4$ are
$\sigma_x\approx 0.10\cs$, 
$\sigma_y\approx 0.15\cs$, and 
$\sigma_z\approx 0.07\cs$, which are too small to be of much significance
for the observable interstellar turbulence in galactic disks.

As mentioned before, symmetric modes in the shearing model B impede the 
midplane penetration of antisymmetric modes, resulting in less inflated 
field lines. The antisymmetric modes have half the azimuthal wavelength
of the symmetric modes, and also shorter radial wavelengths, 
which helps perturbations squeeze 
across the midplane. Nevertheless, the escape of magnetic flux
into high altitude regions is significantly reduced. 
Figure \ref{mass2flux1} shows the variations of $dM/dz$,
$d\Phi_B/dz$, and $dM/d\Phi_B$ for model B.
Compared with
Figure \ref{mass2flux0} for non-sheared model A, Figure \ref{mass2flux1}
for model B illustrates that shear in the Parker instability
results in a configuration where magnetic flux is enhanced in two layers
($0.5 H\simlt |z| \simlt 1.5H$) surrounding the midplane, while the
corona has lower $B$ (higher $\beta$) than the non-sheared case 
(cf, \citealt{shi90}).\footnote{On the other hand, \citet{mil00}
showed that magnetorotational instability in a shearing disk tends to
produce a strongly magnetized corona.}
Apart from the shearing background velocity, small-scale motions are chaotic,
which can facilitate the reconnection of twisted magnetic field lines.
The ``layered'' magnetic profile may in part arise from enhanced
reconnection near the midplane, raising $dM/d\Phi_B$ at 
$|z|/H \simlt 0.5$ (Fig.\ \ref{mass2flux1}$c$).

To summarize, we do not find any firm evidence that shear significantly 
stabilizes the Parker instability. This may be because Parker modes are 
the most unstable at large $\kx$.
By kinematically shifting the $x$-wavenumbers of perturbations at lowest
$k_z$ and $\ky$ which have the largest initial amplitudes -- and are associated
with symmetric modes -- the introduction of shear gives an early
boost to symmetric modes. The more slowly-growing
symmetric modes are later dominated by shorter-wavelength 
antisymmetric modes that grow more strongly (but start from lower
amplitudes). The midplane-centered symmetric modes nevertheless interfere
with the midplane-crossing antisymmetric modes, reducing
the latter's amplification and limiting the inflation of field lines 
toward high latitudes. Except for having slightly more (less)
magnetic support at moderate (low) $|z|$, 
the final state of the shearing model does not differ much overall from 
that of the nonshearing model, consisting of a new vertical equilibrium with 
enhanced central concentration.

\subsection{Gravitational Instability in Shearing Thick Disks}

\subsubsection{Swing Amplification in Unmagnetized Disk}

Self-gravitating evolution of an unmagnetized, thick disk is generally 
very similar to that of a razor-thin disk,
because without magnetic fields the induced motions are almost planar.
One qualitative difference the disk's finite thickness makes is
that it tends to be stabilizing by reducing self-gravity at the midplane
(see \S 3).

As shown in Figure \ref{s25_evol}$b$ as a dotted line, the initial evolution 
of density in model C is governed by nonaxisymmetric swing amplification. 
With $Q=0.7$ and $s_0=25$, hydrodynamic model C is stable to
axisymmetric perturbations (Fig.\ \ref{Qc_s0}).
As wavefronts swing around from leading to trailing, epicyclic motion with
the same rotational sense allows overdense fluid elements to linger
in wave crests. This extended exposure to self-gravity enables 
perturbations to grow until the radial wavenumber becomes large
(\citealt{gol65b,too81}; see also Paper I).
The swing amplification in model C saturates at $t/\torb\approx0.8$, 
producing a state of nonlinearly interacting sheared 
wavelets of relatively high $\kx$. Although the medium continues 
to shear kinematically, nonlinear interactions maintain a 
dominant population of wavelets at moderate $\kx$.
As displayed in the left panel of Figure \ref{swing_col},
for example, the surface density map at $t/\torb=2$ has the most power
at $\kx=7(2\pi/L_x)$.\footnote{Fourier analysis shows that
modes of $\kx > 10(2\pi/L_x$) have negligible power 
throughout the saturation stage, confirming that our simulation results
are not limited by numerical resolution.}
At this time, 
$\langle|\delta v_y|\rangle\approx0.5\langle|v_x|\rangle$ and
$\langle|v_z|\rangle\approx0.04\langle|v_x|\rangle$, indicating
the vertical motions are indeed insignificant compared with the 
planar motions.

Much as we found for two-dimensional models (Paper I), the 
evolution of model C after saturation indicates that nonlinear
interactions repopulate small-$|\kx|$ modes, which then
rejuvenate swing amplification.
This second growth phase yields a few high-density filaments
(middle panel of Fig.\ \ref{swing_col}). The filaments collide 
with each other and become gravitationally unstable, eventually forming 
two gravitationally bound clumps at the end of simulation.
The mass of each clump is $\sim 7\%$ of
the total mass, corresponding to $\sim4\times10^6\Msun$
(right panel of Fig.\ \ref{swing_col});
this is twice the corresponding two-dimensional Jeans mass,
$M_{\rm J}\equiv \cs^4/(G^2\Sigma_0) = 
2\times10^6\Msun (\cs/ 7.0 {\rm\,km\,s^{-1}})^4
(\Sigma_0/65\,\Msun\pc^{-2})^{-1}$,
consistent with the results of Paper I where we found 
the fragments formed in a two-dimensional disk model
with comparable $Q$ and surface density typically have 
$\sim(1/2-2) M_{\rm J}$ (Model H06 in Paper I).

In Paper I, we reported that when $Q$ is sufficiently small, 
razor-thin disks form gravitationally bound clumps via three different
types of secondary instabilities. These include
parallel fragmentation ($Q\simlt0.8$),
gravitationally induced collisions of nonlinear sheared wavelets
($0.8\simlt Q\simlt1.1$), and  rejuvenated swing
amplification ($Q\simgt1.1$ for unmagnetized disks).
Had model C been evolved in the razor-thin approximation,
it would have experienced gravitational runaway via parallel fragmentation 
long before initial swing amplification saturated.
It is remarkable that a thick disk with $Q$ as small as 0.7 
is barely unstable, requiring two successive phases of  
swing amplification before runaway occurs.

\subsubsection{Parker-Swing Instability}

The models of \S4.1.2 and \S4.2.1 represent systems where either gravity
or magnetic fields are entirely turned off, to investigate the physics 
of sheared Parker and thick-disk swing instabilities separately.
Our primary interest here, however, is to understand how magnetic
buoyancy may interact with the self-gravitating instabilities already
known to exist from two-dimensional studies (Paper I).
Model D focuses on the swing-Parker interaction, incorporating self-gravity, 
shear, and thermal-equipartition toroidal magnetic fields. 
Time evolution of the maximum density 
in model D is plotted in Figure \ref{s25_evol}$b$ as a solid line.
Clearly, the initial phase of evolution is dominated by swing amplification,
with very similar behavior to model C (dotted line).
Unlike in models A and B of the Parker instability without self-gravity, 
density in model D begins to grow immediately as $\kx$ increases. 
Embedded magnetic fields reduce the amplification factor in model D 
by 63\% compared to model C.

Swing amplification saturates at about $t/\torb\sim0.8$ when 
$\kx$ becomes sufficiently large.
With embedded azimuthal magnetic fields, however, model D is also 
subject to the Parker instability. 
The shearing wavelets grown from the initial swing amplification 
in fact strengthen the symmetric Parker modes. 
Density in model D increases steadily for $1\simlt t/\torb\simlt 3$; model B 
shows similar behavior at earlier times in Figure \ref{s25_evol}$a$.
At the same time nonlinear interaction of the wavelets -- as discussed
in \S4.2.1 -- keeps the power spectrum centered at moderate $\kx$. 
Figure \ref{all_col} shows surface density snapshots 
at $t/\torb=2$, 3, and 4.1 of model D. 
At $t/\torb=2$, most of the power is concentrated in the mode with
$\kx=9(2\pi/L_x)$ and $\ky=1(2\pi/L_x)$. 
Vertical flows due to the Parker modes couple with in-plane 
(primarily radial) motions, enabling more vigorous nonlinear feedback.
Rejuvenated swing amplification of small-$|\kx|$ modes follows,
and a few dense filaments form (middle of Fig.\ \ref{all_col}). 
They later experience mutual collisions,
ultimately forming two gravitationally bound condensations.
(right of Fig.\ \ref{all_col}). Each clump has 
8\% (left) and 12\% (right) of the total mass, corresponding 
approximately to $9.6\times10^6\Msun$ and $1.4\times10^7\Msun$.

Because of the strong $z$-symmetric perturbations resulting from
the initial swing amplification, model D does not provide
fertile ground for growth of antisymmetric Parker modes.
The latter may grow slightly in high- and middle-altitude regions,
but never become dynamically important.
Consequently, magnetic field lines do not float upward as much
as in the non-self-gravitating model B. 
We draw $dM/dz$, $d\Phi_B/dz$, and $dM/d\Phi_B$ for model D in
Figure \ref{mass2fluxall}, which illustrates that differential mass
and flux at $|z|/H\simgt 1$ are almost unchanged for $t/\torb \simlt 2$. 
Nonlinear interaction of wavelets may cause magnetic field lines to 
reconnect, reducing $d\Phi_B/dz$ near the midplane.

Model D investigated how Parker instability interacts with swing amplification
under conditions of relatively high gaseous surface density such that
the (height-integrated) $Q$ parameter is 0.7 and gas gravity dominates
stellar gravity locally ($s_0=25$).
Regions in which both gas surface density and shear are so high
may, however, be exceptional in contemporary galaxies.
To investigate the effects of the Parker-swing interaction under
other galactic conditions, we also perform low-resolution 
simulations for two cases with lower gaseous surface density.
Both of these models have $s_0=1$; model G has $Q=1.5$ and model I
has $Q=1.0$. For each of these, we also perform simulations
in which self-gravity is turned off (models F and H, respectively)
to provide ``pure Parker'' comparison models. In addition,
we turn off magnetic fields in model J to provide a $Q=1.0$
``pure swing'' counterpart to model I.
The resulting evolutionary histories of maximum density in models
F--J are presented in Figure \ref{s1_evol}. 

When $Q=1.5$ and $s_0=1$, corresponding approximately to average disk
conditions, self-gravity is so weak that the growth of density via 
swing amplification in model G (solid line in Fig.\ \ref{s1_evol}$a$)
is by less than a factor of two.
Since the amplitude of the perturbed density field remains small,
nonlinear feedback from wave interactions do not inhibit the kinematic 
increase of $\kx$.  The medium keeps shearing out, and Parker modes 
slowly emerge during the later part of evolution ($t/\torb\simgt 2.5$).
Model F, in which self-gravity is turned off (dotted line in 
Figure \ref{s1_evol}$a$), has late-stage evolution after saturation
of Parker modes very similar to that of model G. Namely, a new
quasi-static vertical equilibrium, with low-amplitude velocity and
density fluctuations, is achieved. This state is similar to that of
late-stage model B, except that the fluctuation level is lower by
a factor of three. The reason for the differences between the
two non-self-gravitating models (B and F) is that the former has a 
smaller scale height due to a stronger gas self-gravity 
in the initial equilibrium state\footnote{As noted in the beginning
of \S4, our non-self-gravitating (for evolution) models still have 
the initial equilibrium self-gravity that together with external gravity 
balances thermal and magnetic pressure forces in the background state. 
Although the treatment of self-gravity is not self-consistent in model B,
its primary function is as a control for the otherwise identical 
model D, in order to assess the small-scale {\it evolutionary}
effects produced by self-gravity compared to Parker modes alone.}
($H\Omega/\cs$ curves in Fig. \ref{H_s0}), thereby enabling the
Parker instability to grow faster and more strongly (cf.\ eq.\ [\ref{tpar}]). 
This is also why the clear
transition from symmetric to antisymmetric Parker modes that occurs at
$t/\torb\sim1.7$ in model B (Fig.\ \ref{s25_evol}$a$, dotted line)
is not obvious in the density history of model F 
(Fig.\ \ref{s1_evol}$a$, dotted line). 
The conclusive result from the $Q=1.5$, $s_0=1$ model is that
neither the Parker instability nor self-gravity could lead to
significant density condensation
under ``average disk'' conditions that are 
dominated by equipartition-amplitude, large-scale
toroidal magnetic fields.

Models H, I, and J investigate the Parker/swing connection under
conditions of higher disk surface density such that $Q=1.0$;
we set $s_0=1$. The evolutionary histories of $\dmax/\dmid$ are
shown in Figure \ref{s1_evol}$b$. Evidently, comparison of the
development of magnetized (solid line; model I) and unmagnetized
(dashed line; model J) models shows that Parker modes {\it can}
help push a region into self-gravitational runaway when $Q$ is at 
a ``marginal'' level. Model I produces a couple of dense filaments
at $t/\torb=4.8$ and is expected to end in collapse, whereas
the initial swing amplification in model J phase-mixes away within 
$\sim3$ orbits. It is evidently a combination of the extra self-gravity 
produced in ``valleys'' of Parker modes combined with the vorticity-removing 
ability of magnetic fields that tips the balance in allowing a rejuvenated, 
runaway swing phase in model I.
Model H, in which self-gravity is turned off (dotted line),
shows the development of density perturbations for the ``pure Parker''
case, which is quite similar to that of model B (see Fig.\ \ref{s25_evol}$a$),
even though $s_0$ and $Q$ differ. 

In conclusion, the models of this section show that the 
Parker instability does not significantly alter the ability of a large-scale
region of a galactic disk to form gravitationally bound clouds. 
Under ``average disk'' conditions ($Q=1.5$, $s_0=1$, $q=1$), 
the disk turns out to be 
extremely stable, with only mildly fluctuating density in the final state. 
Both magnetized and unmagnetized disks can form bound clouds 
for the high gas surface-density conditions ($Q=0.7$, $s_0=25$)
that represent spiral arm regions (except using ``normal'' shear).
For intermediate surface density ($Q=1.0$),
only magnetized disks form bound clouds. 
While we have not pinpointed a threshold $Q$ value for combined 
Parker-swing nonlinear runaway, models G (see Fig.\ \ref{s1_evol}$a$)
and I (see Fig.\ \ref{s1_evol}$b$) taken together bracket its
value between 1.5 and 1.0 (when $s_0=1$).
Even when bound clouds are created, the Parker instability plays
only an auxiliary role by helping secondary swing amplification;
the formation time scale and spacing of clouds 
are quite different from what the linear-theory Parker analysis predicts. 

\subsection{Magneto-Jeans Instability}

The final issue we address in this paper is the development of the
magneto-Jeans instability (MJI) in a three-dimensional disk. 
We previously studied MJI under uniform weak-shear conditions (Paper I),
and under self-consistent spiral arm conditions (Paper II), both in
a height-integrated two-dimensional idealization. 
While the latter varying shear and varying density approach is 
more realistic, in this first three-dimensional study we adopt the 
simplified uniform model, which is adequate to capture the
basic physical processes: model E has the same 
high-gas-density-parameter set ($Q=0.7$, $s_0=25$) used in model D, but
sets $q=0$ to represent better the reduced (or negative) shear that
prevails within spiral arms (see Paper II).

Physically, MJI is a form of nonaxisymmetric Jeans instability 
in rotating disks containing embedded azimuthal magnetic fields.
Transverse (Alfv\'enic) magnetic perturbations produce radial
tension forces that resist the stabilizing effect
of the Coriolis force; magnetic transfer of angular momentum allows 
secular contraction instead of epicyclic motion
(\citealt{lyn66,elm87}; Paper I).
Although MJI is similar to the Parker instability in that both rely on 
background azimuthal magnetic fields and nonaxisymmetric perturbations,
they are otherwise completely different.
The condition of low shear is essential for 
MJI, because magnetosonic waves begin to quench the Jeans instability
when kinematic shear increases $\kx$; the Parker instability, in contrast,
prefers high $\kx$.
Vertical motions are key in the Parker instability, whereas MJI
involves primarily planar motions.
Since both have been proposed as possible ways to trigger formation of
giant clouds in spiral arms, it is interesting to investigate 
nonlinear evolution when both instabilities can occur. 

Based on model E, we conclude
that MJI, if present, always dominates the Parker instability. 
Although equation (\ref{tpar}) suggests the Parker instability
can have a high initial growth rate, its development never produces 
highly nonlinear near-midplane density perturbations (see solid
line in Fig.\ \ref{s25_evol}$a$).
As the dashed line in Figure \ref{s25_evol}$b$ shows, 
the density in model E ($Q=0.7$) starts to grow immediately and this
growth never saturates, just as we found for MJI in Paper I.
We have also performed low-resolution simulations for MJI in
nonshearing disks with $Q=1.0$ and 1.5 (not listed in Table \ref{tbl-1}), 
and confirmed that MJI can drive them
into gravitational runway. 
The time scale for bound cloud formation becomes longer as $Q$ increases, 
so that disks with $Q=1.0$ and 1.5 form dense clumps within 1 and 2 orbits,
respectively, while model E with $Q=0.7$ takes less than one orbital time
to become fully unstable (Fig.\ \ref{s25_evol}$b$).

In Figure \ref{mji_vol}, we plot perspective views of isodensity
surfaces and magnetic field lines, together with midplane distributions
of density, velocity, and magnetic fields, for model E at $t/\torb=0.5$.
As the perspective view shows, the magnetic fields run almost parallel 
to the $x$--$y$ plane near $z=0$, as do the velocity vectors even during
highly nonlinear stages. 
Centered at $(x/L_x, y/L_y, z/L_z)=(-0.10, -0.19, 0)$ and $(0.14, 0.05, 0)$,
two condensations are produced within one orbital time;
these each contain 27\% and 20\% of the
total mass, corresponding roughly to $\sim3\times 10^7\Msun$. 

As discussed in \S 3.2 for axisymmetric gravitational instabilities, 
the dilution of in-plane self-gravity due to the finite scale height also 
affects the most unstable scale lengths and growth rates of MJI.
The simple replacement $\Sigma_0k\rightarrow 2\dmid Hk/(1+Hk)$ is 
sufficient to take into account this geometrical dilution effect in
an approximate fashion.
Utilizing equation (23) of Paper I (see also \citealt{lyn66}), 
the instantaneous dispersion relation for MJI in thick disks becomes
\begin{equation}\label{mji_disp}
0 = \omega^4 - \left[\kappa^2 + (\cs^2+\vA^2)k^2 - 4\pi G\dmid
\frac{kH}{1+kH}\right]\omega^2 
+\vA^2\ky^2\left(\cs^2k^2 - 4\pi G\dmid\frac{kH}{1+kH}\right),
\end{equation}
where $k^2=\kx(t)^2+\ky^2$ and $\kx(t)=\kx(0)+q\Omega\ky t$. 
Perturbations require $\ky\neq0$ 
(nonaxisymmetric) and $kH_\infty(1+kH)< 2$, or 
$k < k_c \equiv (\sqrt{1+8H/H_\infty} -1)/2H$ to be unstable to MJI
(see eq.\ [\ref{H_inf}] for the definition of $H_\infty$).
Note that the critical wavenumber $k_c$ of MJI is independent
of magnetic pressure (except through $H$) and rotation, so that
MJI operates at smaller scales along the magnetic field 
than axisymmetric gravitational instability.
Since $H/H_\infty\sim1$ for $s_0\simgt 1$ (Fig.\ \ref{H_s0}), 
the condition for MJI is roughly $kH \simlt 1$, which is similar 
to the instability criterion for the Parker modes.\footnote{In investigating
the formation of filamentary structures in the ISM,
\citet{cho00} studied disk models that are similar to
model E, that is, magnetized, uniformly-rotating, and self-gravitating.
They identified the structure forming mechanism with the
Parker-Jeans instability, but we believe that while the 
Parker mechanism may contribute during the linear growth phase,
the primary driver of nonlinear structure formation is the MJI.}

Using equation (\ref{mji_disp}) and the results of \S3.1,
we obtain the peak ($\kx=0$) growth rates for  MJI for model E 
($Q=0.7$, $s_0=25$ and $\beta=1$) as a function of $\ky$, and plot 
them in Figure \ref{mji_growth}.
For comparison, we also plot 
the growth rates for the corresponding razor-thin disk.
The maximum growth rate for a thick disk is $\sim1.12\Omega$
at $k_{y,{\rm max}} = 0.74/H$, so $\lambda_{y,{\rm max}}\sim8H$.
Since $L_y\sim17 H$ for our box, this would predict $\lambda_{y,{\rm max}}/L_y
\sim 0.5$, i.e., two $x$-aligned filaments.
This is indeed consistent with the density
structure in Fig.\ \ref{mji_vol}, although there is also a small contribution
from the $\lambda_{y,{\rm max}}=L_y$ mode that has higher power from
our initial conditions.
Note that the most unstable $\lambda_y$ of MJI is in fact the same as 
that of the dominant Parker mode that grows in non-sheared model A,
although that mode is midplane-antisymmetric in density.
For both $x$- and $y$-scales comparable to $\lambda_{y,{\rm max}}$,
the corresponding mass is $M=\Sigma_0(2\pi/k_{y,{\rm max}})^2
\approx 3\times 10^7 \Msun (\Sigma_0/65\,\Msun\,{\rm pc^{-2}})
(H/80\,{\rm pc})^2$, consistent with the masses of clumps formed 
in model E.\footnote{Because $H$ varies inversely with $\Sigma$
from eq.\ (\ref{H_inf}), the condensation mass would be 
proportional to $\Sigma^{-1}$.}
From Figure \ref{mji_growth}, we note that a razor-thin disk would
characteristically respond to MJI
twice as fast, at half the length scale, of a thick disk.
Consequently, the characteristic Jeans mass is about four times bigger
in a thick disk than the thin-disk prediction.
We note that this result is consistent with simple prediction 
for an unmagnetized disk using the vertical equilibrium condition
of equation (\ref{H_inf}): the central density is
$\pi G\Sigma_0^2/2\cs^2$, and with the three-dimensional Jeans
wavelength $\lambda_{\rm J,3D} = \cs (\pi/G\rho)^{1/2}$,
the nominal three-dimensional Jeans mass would be
$M_{\rm J,3D} = \pi\sqrt{2}\cs^4/(G^2\Sigma_0)$. With
$\lambda_{\rm J,2D}=\cs^2/G\Sigma_0$, the 
two-dimensional Jeans mass is
$M_{\rm J,2D} = \cs^4/(G^2\Sigma_0)\simeq 0.23M_{\rm J,3D}$.

\section{Conclusions}

\subsection{Summary}

A variety of evidence supports the idea that giant, star-forming clouds
form as the result of large-scale collective effects -- i.e., 
dynamical instabilities. Candidate mechanisms that have been proposed
on the basis of linear-theory analysis include the swing amplifier
\citep{gol65b,jul66}, magneto-Jeans instabilities 
(\citealt{lyn66,elm87}; Paper I), 
and the Parker instability \citep{par66}. 
The first two of these involve primarily in-plane motions and are 
fundamentally driven by self-gravity, with either shear or magnetic 
tension acting to neutralize the stabilizing tendency of the Coriolis force. 
For the third mechanism, vertical motion, driven by magnetic
buoyancy, is crucial. 

Based on linear-theory growth rates and characteristic spatial scales, 
all of the above mechanisms could in principle be 
dynamically important under various galactic conditions.
In practice, because giant clouds represent highly overdense
regions compared to mean ISM properties, it is important to 
understand how the candidate instability mechanisms develop
in the {\it nonlinear} regime: i.e., whether and/or how
an instability saturates. Addressing these questions requires
direct numerical simulations to evolve the fluid
dynamics equations into the nonlinear regime.

In previous work, we have investigated the nonlinear development
of the swing and magneto-Jeans mechanisms in a thin-disk
(two-dimensional) approximation (Papers I, II). 
Here, we extend those studies by allowing for fully three-dimensional
dynamics, which also enables us to investigate the nonlinear
development of the Parker instability allowing for rotation,
shear, and self-gravity. Our primary goals were to compare
dynamical development, and especially ``final-state'' 
outcomes, of model disks under various conditions.
By controlling for different physical effects (i.e., turning
gravity, rotation, shear, and magnetic fields ``off'' and ``on''),
and exploring parameter space, we were able to assess the
potential consequences for bound cloud formation of the
various instabilities alone and in combination.

The numerical models we adopt are radially localized, vertically stratified, 
shearing, isothermal disks with embedded magnetic field lines (see \S2.1). 
The initial rate of shear in the background
flow is measured by $q\equiv -d\ln\Omega/d\ln R$, and this overall
shear gradient is maintained by enforcing sheared-periodic radial
boundary conditions; we study cases with $q=1$ and 0.
The initial magnetic field is assumed to
be azimuthal and its vertical stratification is determined such that
the Alfv\'en speed $\vA$ is spatially constant. The density and magnetic
field strength in our model disks are characterized by 
the Toomre parameter $Q$ (see eq.\ [\ref{Toomre_Q}]) and the 
plasma parameter $\beta$ (see eq.\ [\ref{beta}]), respectively. 
Magnetic fields are either initiated at an equipartition level
($\vA/\cs=1$)\footnote{Thus magnetorotational
instability cannot grow; see below.}, or set to zero.
We set up initial magnetohydrostatic equilibria with
thermal and magnetic pressure gradients balancing gaseous self-gravity 
and external gravity from a stellar component. 
We adopt a linear model for the external (stellar-disk) gravity 
(see eq.\ [\ref{extg}]) and parameterize its magnitude using 
$s_0$ (see eq.\ [\ref{s0}]),
which corresponds roughly to the ratio of the gas disk's vertical gravity 
to the external gravity at one scale height. 
The fiducial sets of parameters our model disks adopt are 
$Q=1.5$, $s_0=1$ for the ``average disk'', and 
$Q=0.7$, $s_0=25$ for highly compressed spiral arm regions;
we also perform additional simulations with $Q=1$ and $s_0=1$.

Our simulation boxes are 8 scale-heights ($H$) thick, with area
either $17H\times17H$ or $25H\times25H$ in the $x$--$y$ plane.
The relation between dimensionless simulation parameters and 
physical scales is explained in \S2.2. For all our simulations,
we introduce a spectrum of low-amplitude random perturbations,
and numerically integrate the ideal MHD equations up
to four or five orbital periods. In \S3, 
we first constructed the density profiles for vertical static 
equilibria and calculated the variation of disk thickness as 
a function of $s_0$ (\S3.1), and then studied axisymmetric
gravitational instability of these thick disk models (\S3.2). 
Finite disk thickness reduces the critical value of 
the Toomre stability parameter relative to the zero-thickness limit,
and also increases the spatial scale required for instability. 
We found that for $s_0=1$, $Q_c\sim0.75$, 0.72, and 0.57 
for $\beta=\infty$, 10, and 1 cases, respectively,
suggesting that the average disk is highly stable to axisymmetric
perturbations. We also showed that a simple modification of the 
axisymmetric dispersion relation to account for finite thickness 
(eq.\ [\ref{disp}]) yields excellent agreement with the results of 
simulations for the value of $Q_c$.

In \S4, we present nonlinear evolution of fully three-dimensional
disk models, first considering the Parker instability in isolation (\S4.1),
then investigating evolution of the swing and Parker-swing mechanisms (\S4.2),
and finally considering how self-gravitating perturbations grow under
low shear conditions (\S4.3).

The main results drawn from these three-dimensional numerical simulations 
can be summarized as follows:

1. In the absence of self-gravity, the dominant growing Parker mode under 
uniform rotation is found to be antisymmetric in density with respect to
$z=0$ and tailing with radial wavelength $\lambda_x = 1.9H$ and 
azimuthal wavelength $\lambda_y=8.5H$, and has a growth rate $\sim0.4\cs/H$.
Density fluctuations grow as over-dense regions slide toward the midplane
along undulating magnetic field lines and unloaded portions of the 
field buoyantly rise, but growth slows as nonlinear saturation sets in.
The maximum density at saturation is less than twice the initial midplane 
density. After saturation of the dominant Parker mode, nonlinearly
developed small-scale modes interact with each other, 
dispersing density structures characteristic of the Parker instability.
Escape of magnetic flux across the vertical boundaries and likely
reconnection of the twisted magnetic fields prompted by small-scale motions
raise the mass-to-flux ratio near the midplane.
The system progressively moves to a new global vertical equilibrium in 
which the magnetic field is nearly uniform, hence magnetic pressure support 
against gravity is negligible. In the saturated state,
small scale fluctuating velocities are still present, but their 
density-weighted dispersions are less than 10\% of the isothermal speed speed.

2. When a disk is subject to differential rotation, midplane symmetric 
(in density) Parker modes with $\lambda_y=17H$ are the first to dominate the
evolution. This initial dominance is explained by the large initial
amplitude of longer-wavelength perturbations, combined with the ability
of shear to increase $\kx$ to the regime where Parker instabilities 
are more efficient.
Smaller-$\lambda_y$ antisymmetric modes start with lower amplitudes than 
the symmetric modes but grow more rapidly, moving gradually downwards 
from high altitudes to interact with the symmetric mode.
The mode interaction reduces magnetic field line inflation and loss 
compared to the non-shearing case.
The final state of the non-self-gravitating shearing disk model is similar 
to that of the non-shearing disk, consisting of an overall vertical 
equilibrium with small-scale, low-amplitude fluctuations in density and 
velocity fields. 

3. When we include both self-gravity and uniform shear in our models,
the nonlinear evolution of structure is principally dominated by the
swing amplification mechanism, although Parker modes play a role
in seeding structure.
Low-$Q$ ($Q=0.7$) disk models end in gravitational runaway,
while high-$Q$ ($Q=1.5$) disk models remain extremely stable. 
In both high and low $Q$ limits, the ultimate outcome is decided 
independent of the presence or absence of magnetic fields. 
When $Q$ is marginal ($Q=1$), on the other hand, we find that
only magnetized models form bound condensations, suggesting that
Parker instability may play a supplementary role in destabilization.
The Parker instability raises the level of density fluctuation and
may help channel power in the saturated state into small wavenumber modes.
It is the ``rejuvenated'' swing amplification (see Paper I) that 
ultimately drives the system to form dense condensations, however.
For cases in which bound clouds do form, the sizes and spacings are 
not characteristic of Parker modes, but comparable to the results 
from thin disk models (with typical masses $M\sim (2-3)$ times
the thin disk Jeans mass).

4. Weakly shearing thick disks, like thin disks, are susceptible to the
magneto-Jeans instability (MJI), in which tension forces from 
azimuthal magnetic fields vitiate the stabilizing Coriolis force.
The azimuthal wavelength criterion for MJI in finite-thickness disks 
is similar to that of the Parker instability ($\ky H\simlt1$), 
and modal growth rates are also similar.
Our nonlinear simulations show, however, that MJI under weak-shear (spiral arm)
conditions is ultimately much more violent than the Parker instability, 
and can form dense condensations within one orbital time.
This is because the Parker instability is self-limiting, whereas 
the MJI is a runaway process. The mass of dense condensations amounts 
typically to the thick-disk Jeans mass ($\sim$ 4 times the
thin-disk Jeans mass).

\subsection{Discussion}

The numerical simulations in this paper suggest that formation of bound 
clouds cannot occur as a direct consequence of Parker instability,
although Parker instability may play a role in seeding swing amplification
in high-shear environments or MJI in regions of weak shear. 
There are several reasons why the Parker instability is ineffective
at producing large, dense clouds.
First, although the growth time of the Parker instability from linear-theory
analyses is a few times $10^7$ yrs (cf, \citealt{kimj98_1}), 
the process does not proceed in a runaway fashion, but is
stabilized at a relatively modest density by magnetic tension forces
(e.g., \citealt{mou74}).
Second, while the undular Parker instability prefers wavelengths
in the azimuthal direction $\lambda_y\sim 2\pi H$ (i.e., $\ky H\sim 1$ 
in rough agreement with observed spacing between molecular complexes),
there is no similar preferred scale in the radial direction
(e.g., \citealt{par67,giz93,kimj98_2}). 
Growth rates increase slowly with $\kx$, and are nearly uniform 
for $\kx H \simgt 5$. 
Indeed, even though the initial perturbation amplitudes in our simulations
increase toward large scale, we find that small-$\lambda_x$ modes dominate
the evolution. The results is long filamentary structures with 
a radial spacing about 10 times smaller than the azimuthal
spacing in surface density maps during the initial growing phase
(see Fig.\ \ref{col1}).\footnote{The corresponding radial wavelengths 
are one or two times the disk scale height for symmetric and 
antisymmetric structures, respectively, possibly indicating nonlinear
selection of these modes.} 
Because the perturbed self-gravitational potential 
$\delta\Phi_s\propto - \delta\rho /k^2$ is increasingly small for large $\kx$, 
effects of self-gravity on development of the Parker-Jeans instability are 
weak for the dominant nonlinear modes. 

Third, while two-dimensional simulations for the Parker instability
inarguably reach \citet{mou74}-type final static equilibrium configurations 
(cf, \citealt{san00}), three-dimensional simulations 
(\citealt{kimj98_2,kimj00}, and this work) indicate
that such equilibrium configurations are transient at best,
evolving rapidly into a state where density is relatively smooth
in the horizontal direction. 
It is unclear whether this transition is initiated by an instability
of the static undulating configurations, as claimed by 
\citet{lac80}, by active magnetic reconnection near
the midplane caused by small-scale chaotic motions, or
simply by comparable nonlinear saturation levels of many high-$\kx$ modes.
When shear is present, kinematically driven phase mixing contributes
to washing out large-scale radial structures.
In any case, the three-dimensional simulations show that it is
almost impossible to find any signature of the undular Parker instability
in the later stages of evolution.

The result that Parker instability cannot {\it directly} account for GMC 
formation has previously been emphasized by other authors.
\citet{kimj01}, for example, find only a maximum factor of 1.5
enhancement in surface density in their three-dimensional 
non-self-gravitating simulations with solid-body rotation, 
and conclude that Parker mode {\it alone}
cannot be responsible for forming molecular clouds 
(see also e.g., \citealt{bas97,kimj98_2,kimj00,san00}). 
These authors speculate on the possibility of the Parker instability 
coupled with other physical processes such as self-gravity, 
differential rotation, radiative cooling, and spiral arm potentials.
We show in the present paper that self-gravity and galactic differential 
rotation are not much help. 
Therefore, we argue that {\it the Parker instability, even
combined with gaseous self-gravity and differential rotation,
cannot be regarded as the primary formation mechanism for giant
molecular clouds at least in galactic disks at large}.
While effects on the Parker instability of stronger self-gravity and 
rapidly varying shear inside spiral arms have yet to be studied 
in detail, we do not anticipate significant differences in the
main conclusions. Although the width of a spiral arm might
help in promoting Parker modes of similar $\lambda_x$ over
shorter-wavelength competitors (cf., \citealt{fra02}), it is not
yet clear how effective this can be if the arm transit time is 
comparable to the growth time.

While larger or smaller shear rates do not seem to alter the 
basic character of Parker instabilities appreciably, the shear rate
is crucial for selecting which type of (in-plane) gravitating
amplifier can operate. In a weakly shearing regions, swing is absent,
but the MJI is very powerful at initiating self-gravitational condensations.
In Paper I, we showed MJI is a potential means to prompt observed 
active star formation near the central parts of galaxies 
where rotation curves are nearly rigid-body (see also \citealt{elm87}).
In Paper II, we argued that observed spur structures jutting radially 
outwards from spiral arms can be understood as a direct consequence
of the MJI occurring within the arms. 
A spiral arm potential has three important effects on the gaseous 
medium. (1) By compressing gas, the Jeans length -- which determines
the preferred MJI scale along the mean magnetic field -- is reduced;
the growth time for self-gravitating modes also drops. 
(2) The spatially-varying density distribution
causes a gradient in the shear profile perpendicular to the arm;
since transiting gas first experiences reversed shear and then returns
to normal shear, the spatially-varying profile tends to keep
radial wavenumbers small within the arm, which enhances the efficiency 
of MJI. (3) It reduces the radial scale length for
mass collection by MJI to at most  the arm width, so that the masses of 
clumps produced are limited. 

Compared to our previous results from razor-thin disks, the weaker 
self-gravity in three dimensions could imply somewhat more massive 
($\sim10^7\,\Msun$) clouds would tend to
form from MJI under realistic conditions.
On the other hand, the weaker self-gravity may allow stronger
spiral potential perturbations that give a higher density
contrast between arm and interarm regions in stable 
quasi-axisymmetric spiral shock profiles (cf.\ Paper II).
The Jeans mass at the spiral arm density peak could be correspondingly 
smaller, producing lower mass clouds in three-dimensional spiral
arm regions.
Since the marginal wavenumber of MJI parallel to $\mathbf{B}$ is about the same
as that of the Parker instability in spiral arm conditions (see \S4.3),
one might expect some cooperation between MJI and the Parker instability 
inside spiral arms. The present models under uniform in-plane 
background conditions do not
point to significant modifications of the nonlinear MJI by Parker modes. 
Effects from Parker-MJI interactions could be more prominent with
a realistic spiral arm model (cf. \citealt{mar98}), however, and it would be 
particularly interesting to learn whether the final mass scale
changes markedly compared to predictions from uniform disk models.

By focusing on cases with relatively strong ($\beta$=1) equilibrium 
azimuthal magnetic fields in our MHD models, we have concentrated solely on 
effects of the Parker instability and MJI in the present work.
An additional dynamical process prevalent in magnetized disk systems is the
magnetorotational instability (MRI; \citealt{bal91,bal98}).
Because MRIs are most vigorous in their small-scale
growth when the thermal pressure is large compared to magnetic
pressure ($\beta\gg 1$), they have been studied primarily in the 
context of accretion disks. However, \citet{sel99} have
argued that the MRI may also be important in producing turbulence in
galactic disks where the effective value of $\beta$ is close to unity, 
citing as evidence the significant nonthermal velocity dispersions 
($\sim 6\,{\rm km\, s^{-1}}$) present in outer \ion{H}{1} disks 
with low local star formation rates and apparent gravitational stability. 
It is of great interest to investigate whether 
coupling between the MRI and self-gravitating modes under galactic 
conditions could represent an important new mechanism for forming 
large-scale ISM condensations.

\acknowledgements
We are grateful to an anonymous referee and N.\ Turner 
for thoughtful comments and to
S.\ White, T.\ Sano, and J.\ Lee for help
in using IDL to make three-dimensional visualization.
This work was supported by NASA grant NAG 59167.

\clearpage
\begin{deluxetable}{ccccccccl}
\tabletypesize{\footnotesize}
\rotate
\tablecaption{Parameters of Three-Dimensional Simulations. \label{tbl-1}}
\tablewidth{0pt}
\tablehead{
\colhead{\begin{tabular}{c} Model        \\ (1) \end{tabular} } &
\colhead{\begin{tabular}{c} $Q$          \\ (2) \end{tabular} } &
\colhead{\begin{tabular}{c} $s_0$        \\ (3) \end{tabular} } &
\colhead{\begin{tabular}{c} $\beta$      \\ (4) \end{tabular} } &
\colhead{\begin{tabular}{c} $q$          \\ (5) \end{tabular} } &
\colhead{\begin{tabular}{c} Gravity      \\ (6) \end{tabular} } &
\colhead{\begin{tabular}{c} Grid         \\ (7) \end{tabular} } &
\colhead{\begin{tabular}{c} Size         \\ (8) \end{tabular} } &
\colhead{\begin{tabular}{c} Relevant Physics
                                         \\ (9)\end{tabular} }
}
\startdata
A & 0.7 & 25 & 1      & 0 & no  & $256\times128\times128$ & $17\times17\times8$ & Parker+Rotation \\
B & 0.7 & 25 & 1      & 1 & no  & $256\times128\times128$ & $17\times17\times8$ & Parker+Shear    \\
C & 0.7 & 25 &$\infty$& 1 & yes & $256\times128\times128$ & $17\times17\times8$ & Swing           \\
D & 0.7 & 25 & 1      & 1 & yes & $256\times128\times128$ & $17\times17\times8$ & Parker+Shear+Gravity \\
E & 0.7 & 25 & 1      & 0 & yes & $128\times128\times128$ & $17\times17\times8$ & MJI             \\
  &     &    &        &   &     &                         &    &                 \\
F & 1.5 & 1  & 1      & 1 & no  & $64\times64\times64$    & $25\times25\times8$ & Parker+Shear    \\
G & 1.5 & 1  & 1      & 1 & yes & $64\times64\times64$    & $25\times25\times8$ & Parker+Shear+Gravity  \\
  &     &    &        &   &     &                         &    &                 \\
H & 1.0 & 1  & 1      & 1 & no  & $64\times64\times64$    & $25\times25\times8$
& Parker+Shear     \\
I & 1.0 & 1  & 1      & 1 & yes & $64\times64\times64$    & $25\times25\times8$
& Parker+Shear+Gravity  \\
J & 1.0 & 1  &$\infty$& 1 & yes & $64\times64\times64$    & $25\times25\times8$ & Swing   \\
\enddata

\tablecomments{See \S4 for explanation.} 
\end{deluxetable}

%1
\clearpage
\begin{figure}
\epsscale{1.}
%\plotone{H_s0.ps}
\caption{Variation of the vertical scale height $H$ of the gas distribution 
in a magnetohydrostatic equilibrium with varying
gaseous-to-stellar surface density ($\propto s_0^{1/2}$). 
The $H/H_\infty$ curves show the thickness relative to that of a purely
self-gravitating gas disk, while
the $H\Omega/\cs$ curves show the thickness relative to a fixed
fiducial value. Strong self- or external gravity leads to 
a smaller scale height. Magnetic pressure makes a disk thicker.
\label{H_s0}}
\end{figure}

%2
%\clearpage
\begin{figure}
\epsscale{1.}
%\plotone{ext_grav.ps}
%\vspace{-3.cm}
\caption{Height dependence of driving force $\geff$ (solid lines) of the 
Parker instability for two model galactic disks.
Dotted and short-dashed lines correspond to self-gravity and external gravity, 
respectively. 
({\it a}) Under average disk conditions ($Q=1.5$ and $s_0=1$), 
gaseous self-gravity is comparable to the external gravity up 
to $z/H\sim1$, beyond which the latter dominates. For comparison, 
we plot the external gravity models adopted by \citet{giz93} and 
\citet{kimj98_1} using dot-dashed and long-dashed lines, respectively. 
({\it b}) For strongly-compressed gas in a spiral arm ($Q=0.7$ and $s_0=25$),
self-gravity is much more important than external gravity.
\label{ext_grav}}
\end{figure}

%3
%\clearpage
\begin{figure}
\epsscale{1.}
%\plotone{Qc_s0.ps}
\caption{({\it a}) The most unstable wavenumber $\kmax$ and 
({\it b}) critical $Q_c$ for axisymmetric gravitational 
instability in a finite-thickness disk. Various symbols (circles for 
$\beta=\infty$, triangles for $\beta=10$, and squares for $\beta=1$)
plotted in ({\it b}) are 
obtained by numerical simulations, showing  excellent agreement 
with the analytic estimates. See text for details.
\label{Qc_s0}}
\end{figure}

%4
%\clearpage
\begin{figure}
\epsscale{1.}
%\plotone{s25_evol.ps}
\caption{Time evolution of maximum density in models A -- E with $Q=0.7$ 
and $s_0=25$. ({\it a}) Models without self-gravity show
only moderate increase of density, never developing highly 
nonlinear perturbations. ({\it b}) With self-gravity included, the
disk models eventually become gravitationally unstable, forming
bound clumps at the end. 
\label{s25_evol}}
\end{figure}

%5
%\clearpage
\begin{figure}
\epsscale{1.}
%\plotone{parker3d.ps}
\caption{Perturbed density (color scale) in three different planes
($x=-0.5L_x$, $y=-0.5L_y$, and $z=0.12L_z$)
and selected magnetic field lines for ({\it left}) model A and ({\it right}) 
model B at $t/\torb=1.6$. 
Only the portions of magnetic fields that start from
$x=-0.5L_x$ and $z=0.12L_z$ and rise above the $z=0.12L_z$ plane
are shown.
The midplane ($z=0$) is marked by thin line in each box and colorbars
label $\delta\rho$ in units of the initial central density. 
\label{parker3d}}
\end{figure}

%6
%\clearpage
\begin{figure}
%\vspace{4cm}
\epsscale{1.}
%\plotone{parker_q0_col.ps}
%\vspace{-5cm}
\caption{Snapshots from model A of the Parker instability with uniform 
rotation.
({\it a}) Total surface density map ($x$--$y$ plane) at $t/\torb=2$.
({\it b}) Density structure at the midplane ($z=0$) at $t/\torb=2$.
({\it c}) Square section in ({\it b}) is enlarged to show density and 
horizontal velocity vectors in detail. The arrow outside the box shows
the sound speed for comparison. ({\it d}) Surface density map at $t/\torb=3$.
Gray scales are linear in units of $\Sigma_0$ or $\dmid$.
\label{col1}}
\end{figure}

%7
%\clearpage
\begin{figure}
\epsscale{1.}
%\plotone{mass2fluxq0.ps}
\caption{({\it a}) Differential mass $dM/dz$ and  
({\it b}) magnetic flux $d\Phi_B/dz$, averaged aver 
the $y$-axis, and ({\it c}) differential mass-to-flux ratio 
$dM/d\Phi_B$ of the Parker instability simulation with uniform
rotation (model A) are drawn as functions of the vertical height $z$. 
\label{mass2flux0}}
\end{figure}

%8
%\clearpage
\begin{figure}
\epsscale{1.}
%\plotone{q1_col.ps}
\caption{Surface density maps projected on the $x$--$y$ plane 
of the simulation for the 
Parker instability including shear (model B) at $t/\torb=1$, 2, and 3
from left to right. Gray scale bars label $\Sigma/\Sigma_0$.
\label{q1_col}}
\end{figure}

%9
%\clearpage
\begin{figure}
\epsscale{1.}
%\plotone{mass2fluxq1.ps}
\caption{Same as Figure \ref{mass2flux0} except for model B that
contains shear.
\label{mass2flux1}}
\end{figure}

%10
%\clearpage
\begin{figure}
\epsscale{1.}
%\plotone{swing_col.ps}
\caption{Snapshots of surface density of the hydrodynamic model C
at $t/\torb=$2.0, 2.9, and 3.4 from left to right.
Numbers labeling gray-scale bars correspond to $\log \Sigma/\Sigma_0$.
\label{swing_col}}
\end{figure}

%11
%\clearpage
\begin{figure}
\epsscale{1.}
%\plotone{all_col.ps}
\caption{Snapshots of surface density of model D
at $t/\torb=$2.0, 3.3, and 4.1 from left to right.
Numbers labeling gray-scale bars correspond to $\log \Sigma/\Sigma_0$.
\label{all_col}}
\end{figure}

%12
%\clearpage
\begin{figure}
\epsscale{1.}
%\plotone{mass2flux_all.ps}
\caption{({\it a}) Differential mass $dM/dz$ and  
({\it b}) magnetic flux $d\Phi_B/dz$, averaged aver
the $y$-axis, and ({\it c}) differential mass-to-flux ratio
$dM/d\Phi_B$ of model D.
\label{mass2fluxall}}
\end{figure}

%13
%\clearpage
\begin{figure}
\epsscale{1.}
%\plotone{s1_evol.ps}
\caption{Time evolution of maximum density for ({\it a}) models
F and G with $Q=1.5$ and ({\it b}) models
H -- J with $Q=1.0$. In this plot, $G\neq0$ indicates self-gravitating
cases, while non-self-gravitating cases are marked by $G=0$.
\label{s1_evol}}
\end{figure}

%14
%\clearpage
\begin{figure}
\epsscale{1.}
%\plotone{mji_vol.ps}
\caption{{\it Left:} Perspective visualization of isodensity surfaces 
($\rho/\dmid=1.5$) and selected magnetic field lines in model E 
at $t/\torb=0.5$.  
Starting positions of magnetic field lines lie at $y/L_y=-0.5$ and 
$z/L_z=-0.45$, 0, and 0.45 from bottom to top, respectively.
{\it Right:} Density in logarithmic gray scale, 
velocity vectors (arrows), and magnetic field lines (thick solid lines) 
at the midplane in model E at $t/\torb=0.5$. Note that $z$-components
of velocity and magnetic fields are almost zero near the midplane.  
\label{mji_vol}}
\end{figure}

%15
%\clearpage
\begin{figure}
\epsscale{1.}
%\plotone{mji_growth.ps}
\caption{Growth rates of magneto-Jeans instability for 
thick and thin disk models with
$Q=0.7$, $s_0=25$, $\beta=1$, and $\kx=0$. Overestimated
self-gravity in thin disks implies 
larger growth rates and smaller unstable length scales. 
\label{mji_growth}}
\end{figure}

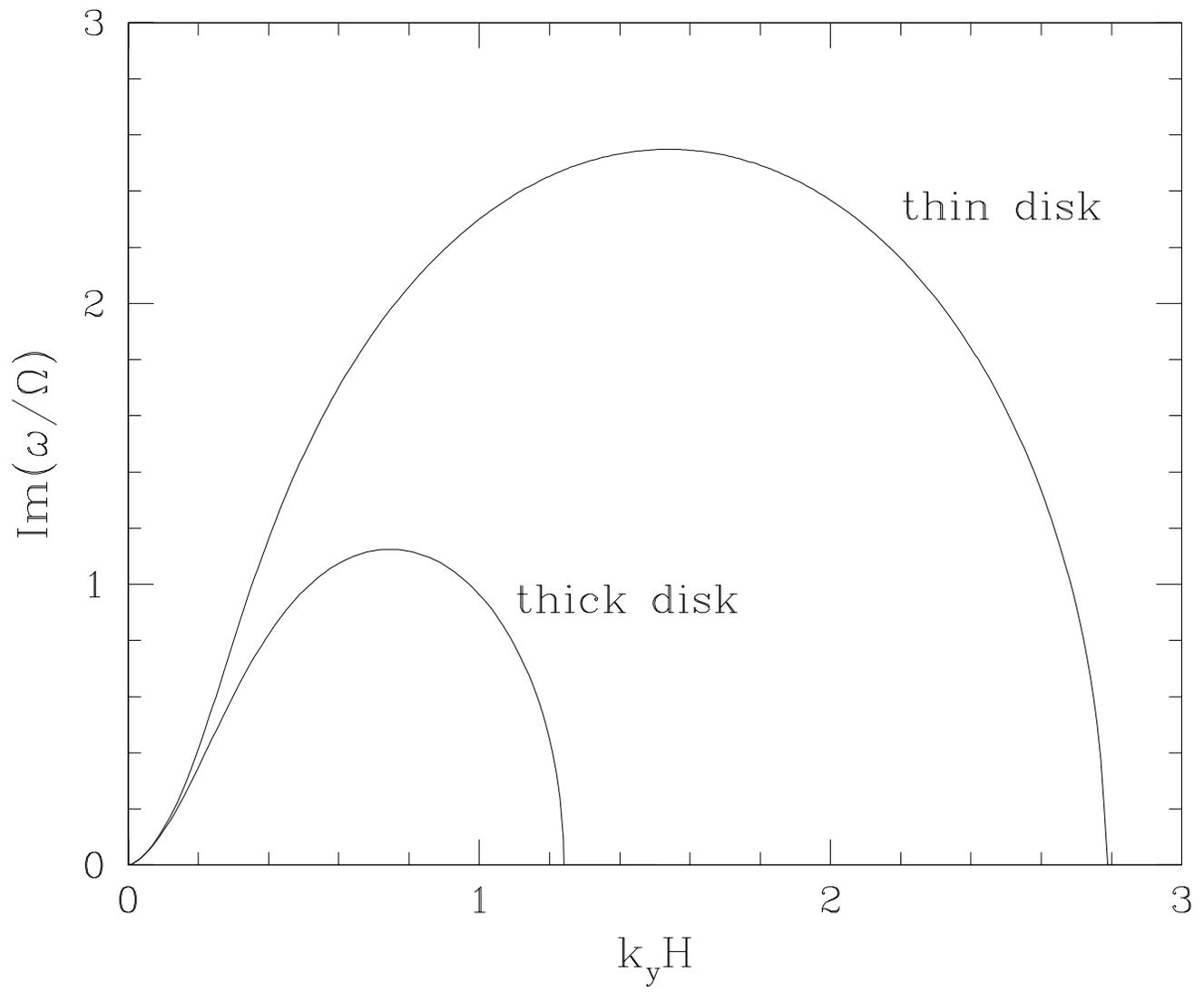
\end{document}